\documentclass[fleqn,usenatbib]{mnras}

\usepackage{newtxtext,newtxmath}

\usepackage[T1]{fontenc}

\DeclareRobustCommand{\VAN}[3]{#2}
\let\VANthebibliography\thebibliography
\def\thebibliography{\DeclareRobustCommand{\VAN}[3]{##3}\VANthebibliography}

\renewcommand{\vec}[1]{ {\bf #1} }

\usepackage[normalem]{ulem}

\usepackage{graphicx}	% Including figure files
\usepackage{amsmath}	% Advanced maths commands
\usepackage{amssymb}	% Extra maths symbols
\newcommand{\vect}[1]{\boldsymbol{#1}}

\title[Supersonic turbulence with high-order DG]{Supersonic turbulence simulations with GPU-based high-order Discontinuous Galerkin hydrodynamics}

\author[M. Cernetic et al.]{%
Miha Cernetic$^{1}$\thanks{E-mail: cernetic@mpa-garching.mpg.de},
Volker Springel$^{1}$,
Thomas Guillet$^{2}$,
and R\"udiger Pakmor$^{1}$
\vspace*{1mm}\\%
$^{1}$Max Planck Institut f\"ur Astrophysik, Karl-Schwarzschild-Straße 1, 85748 Garching bei M\"unchen, Germany\\%
$^{2}$Physics and Astronomy, University of Exeter, Exeter EX4 4QL, UK
}

\date{Accepted 2024 September 11. Received 2024 August 20; in original form 2023 December 20}

\pubyear{2023}

\begin{document}
\label{firstpage}
\pagerange{\pageref{firstpage}--\pageref{lastpage}}
\maketitle

\begin{abstract}
We investigate the numerical performance of a Discontinuous Galerkin (DG) hydrodynamics implementation when applied to the problem of driven, isothermal supersonic turbulence. While the high-order element-based spectral approach of DG is known to efficiently produce accurate results for smooth problems (exponential convergence with expansion order), physical discontinuities in solutions, like shocks, prove challenging and may significantly diminish DG's applicability to practical astrophysical applications. We consider whether DG is able to retain its accuracy and stability for highly supersonic turbulence, characterized by a network of shocks. We find that our new implementation, which regularizes shocks at sub-cell resolution with artificial viscosity, still performs well compared to standard second-order schemes for moderately high Mach number turbulence, provided we also employ an additional projection of the primitive variables onto the polynomial basis to regularize the extrapolated values at cell interfaces. However, the accuracy advantage of DG diminishes significantly in the highly supersonic regime. Nevertheless, in turbulence simulations with a wide dynamic range that start with supersonic Mach numbers and can resolve the sonic point, the low numerical dissipation of DG schemes still proves advantageous in the subsonic regime. Our results thus support the practical applicability of DG schemes for demanding astrophysical problems that involve strong shocks and turbulence, such as star formation in the interstellar medium. We also discuss the substantial computational cost of DG when going to high order, which needs to be weighted against the resulting accuracy gain. For problems containing shocks, this favours the use of comparatively low DG order.
\end{abstract}


%%%%%%%%%%%%%%%%%%%%%%%%%%%%%%%%%%%%%%%%%%%%%%%%%%

%%%%%%%%%%%%%%%%% BODY OF PAPER %%%%%%%%%%%%%%%%%%

\section{Introduction}

Turbulence is a fundamental physical phenomenon that appears universally in fluid flow \cite[e.g.][]{Launder1974,Larson1981,Mellor1982,Kim1987,Menter1994,Frisch1995,Goldreich1995,Balbus1998,Pope2000,Brandenburg2011}, and thus affects many fields of study, including meteorology, engineering, and, of course, astrophysics. For example, there is turbulence in and around the Sun, something that will be further characterized by a recently approved NASA Medium-Class Explorer mission  \citep{Klein2023}. In our Galaxy, the interstellar medium (ISM) is characterized by supersonic turbulent motions that shape the gas distribution and gas kinematics,  and that play a fundamental role in regulating star formation, as first observed by \citet{Larson1981}, with a recent review on the topic by \citet{BallesterosParedes2020}. Multiple recent works used ALMA to study the influence of turbulence on star formation \citep[e.g.][]{Li2020,Gomez2021,Bhadari2023}. The ongoing PASIPHAE~\citep{pasiphae2018} and POSSUM~\citep{possum2021} surveys will soon produce a full tomographic map of the galactic magnetic field, shedding new light on the nature ISM and CGM turbulence. 

The observational interest in ISM turbulence is matched only by the vast number of theoretical investigations. Because turbulence has a commanding influence on the distribution of gas in the ISM, many studies, dating back decades, have looked into this \citep[e.g.][]{scalo_probability_1998, passot_density_1998, ostriker_kinetic_1999, klessen_one-point_2000, wada_numerical_2001, ballesteros-paredes_physical_2002, li_formation_2003, kravtsov_origin_2003, mac_low_distribution_2005,Federrath2009,Federrath2021,SajaySunny2023,Rabatin2023}. Using the results from these studies a series of new star formation recipes were proposed by, e.g., \citet{Kretschmer2020} and \citet{Girma2024}, among others.

Hydrodynamical simulations are a primary tool for the study of such highly non-linear physics. But numerical effects and resolution limitations strongly influence the quality of the obtained hydrodynamical results, motivating a constant search for improvements in numerical schemes and likewise demanding careful validation of new techniques.

In this paper, we investigate the main properties and effects of high-order Discontinuous Galerkin (DG) methods when applied to supersonic turbulence. The DG approach is a general tool of applied mathematics first used to solve the equations of neutron transport by \citet{Reed1973} and then robustly defined in a series of five papers by \citet[][]{Cockburn1988, Cockburn1989,Cockburn1989a,Cockburn1990,Cockburn1998}. DG has been gaining traction as a key method for solving partial different equations, such as the Euler equations of fluid flow used in multiple recent works \citep{Schaal2015,velasco_romero2018,guillet_high_order_2019}.

In a companion paper \citep{subsonic_dg_gpu}, we have presented an implementation of a GPU-accelerated, MPI-parallel DG code for solving the Navier-Stokes equations. We could confirm the high accuracy and computational efficiency of this approach in a variety of test problems, even showing exponential convergence as a function of the employed spectral order. We also demonstrated that shocks and physical discontinuities can be handled by an artificial viscosity field at sub-cell resolution. The width of these continuities is however fundamentally limited by the effective spatial resolution of the scheme, and thus only linearly improves with higher spatial resolution, as is the case with ordinary finite volume schemes. This raises the important question whether the advantages of DG are defeated in problems containing many shocks, a question we seek to address in this paper.

A physical setting where this perhaps can be answered in a particularly succinct way is supersonic, isothermal turbulence. The density probability distribution function (PDF) and the power spectrum of compressible, supersonic turbulence play an important role especially in theories of star formation.
% \citep[e.g.][]{}. 
However, super- and hypersonic turbulence are particularly challenging for Eulerian mesh codes given the extremely high ram pressures, strong shocks, and huge density contrasts that develop in this regime, in addition to regions of nearly vanishing density. This makes it hard to capture the inertial range of supersonic turbulence accurately, even more so than for subsonic turbulence.

In DG methods, the solution inside cells is approximated by smooth, high-order polynomials. It is clear that strong shocks passing through cells may play havoc with such polynomials, causing strong Gibbs-like oscillations, and worse, potentially trigger so wide oscillations that unphysical values of fluid variables occur. To address this, we in this paper introduce a modification of our sub-cell shock capturing scheme -- actually simplifying it considerably compared to our previous approach -- by resorting to a classic von Neumann-Richtmyer viscosity \citep{VonNeumann1950}. In addition, we introduce a novel regularisation of the primitive variables at cell boundaries, which proves critical to stably and accurately evolve high Mach number turbulence with DG at high order.

In this paper, we demonstrate the accuracy of these new implementations by considering  a number of test problems containing strong shocks. We then move on to simulations of driven, isothermal turbulence. We vary the Mach number systematically from low values to Mach numbers beyond ten, comparing at each stage DG with a standard, second-order finite volume method based on piece-wise linear reconstruction. We analyze velocity power spectra, structure functions and density PDFs in order to examine the advantages brought by going to DG as compared to classic finite volume (FV) methods with the same number of cells. We also include an analysis of high dynamic range simulations that can resolve the sonic point.

The paper is structured as follows. First, in Section~\ref{sec:intro_to_DG}, we summarize the DG approach in general and the particular implementation we have developed in our GPU-based code.  Then, in Section~\ref{Sec:method_improvements_for_supersonic} we introduce two method improvements in the form of a new artificial viscosity treatment and a projection method for the primitive fluid variables. Combined, they allow sustained high mach number turbulence simulations with DG.
In Section~\ref{sec:supersonic_driving} we detail our implementation of turbulence driving and our measurements of basic turbulence statistics. Section~\ref{sec:turbulenceruns} presents our main simulation results in the form of a systematic suite of turbulence simulations, from the subsonic to the supersonic regimes. For a specific choice of driving, we extend the dynamic range in Section~\ref{sec:sonicpoint} substantially by going to higher resolution, allowing us to see the transition from supersonic to subsonic turbulence at the sonic point.  We discuss the computational cost of high-order DG in Section~\ref{sec:discussion}, and conclude by summarizing our results in Section~\ref{sec:conclusions}.

\section{Discontinuous Galerkin hydrodynamics}
\label{sec:intro_to_DG}

The Discontinuous Galerkin (DG) approach is a general high-order finite element method for numerically solving partial differential equations \citep[e.g.][]{Cockburn1989}. Here we apply it to the Euler and Navier-Stokes equations for numerical hydrodynamics. Consider the Euler equations
\begin{equation}
    \label{eq:euler_equation}
    \frac{\partial \boldsymbol{u}}{\partial t}+\sum_{\alpha=1}^{d} \frac{\partial \boldsymbol{f}_{\alpha}(\boldsymbol{u})}{\partial x_{\alpha}}=0 ,
\end{equation}
with the state vector $\boldsymbol{u}$ storing the conserved variables of each cell, and the sum $\alpha$ running over their spatial dimensions and $\boldsymbol{f}_{\alpha}(\boldsymbol{u})$ being the analytical flux matrix. The state vector consists of 
\begin{equation}
    \label{eq:state_vector_definition}
    \boldsymbol{u}=\left[\begin{array}{c}
\rho \\
\rho \boldsymbol{v} \\
e \\
\end{array}\right], \quad e=\rho u+\frac{1}{2} \rho \boldsymbol{v}^{2},
\end{equation}
with $u$ being the specific internal thermal energy, while $\rho$ denotes the fluid density, $\boldsymbol{v}$ its velocity, and $e$ its total energy density. To completely describe the gas we also need an equation of state connecting the pressure $p$ with $u$ and $\rho$. For this we employ the ideal gas equation,
\begin{equation}
    p = \rho u \left( \gamma - 1 \right),
    \label{eq:ideal_gas}
\end{equation}
where $\gamma$ is the ratio of specific heats at constant pressure and constant volume, respectively, commonly known as the adiabatic index. The flux matrix $\boldsymbol{f}_{\alpha}(\boldsymbol{u})$, spelled out explicitly in 3D, is given by
\begin{equation}
    \label{eq:flux_matrix}
    \boldsymbol{f}_{1}=\left(\hspace*{-0.2cm}
    \begin{array}{c}
        \rho v_{x} \\
        \rho v_{x}v_{x}+p \\
        \rho v_{x} v_{y} \\
        \rho v_{x} v_{z} \\
        (\rho e+p) v_{x}
    \end{array}\hspace*{-0.2cm}\right), \; \boldsymbol{f}_{2}=\left(\hspace*{-0.2cm}
    \begin{array}{c}
        \rho v_{y} \\
        \rho v_{x} v_{y} \\
        \rho v_{y}v_{y} +p \\
        \rho v_{y} v_{z} \\
        (\rho e+p) v_{y}
    \end{array}\hspace*{-0.2cm}\right), \; \boldsymbol{f}_{3}=\left(\hspace*{-0.2cm}
    \begin{array}{c}
    \rho v_{z} \\
    \rho v_{x} v_{z} \\
    \rho v_{y} v_{z} \\
    \rho v_{z}v_{z}+p \\
    (\rho e+p) v_{z}
    \end{array}\hspace*{-0.2cm}\right).
\end{equation}

The key starting point of the DG method is to approximate the solution of the Euler equations~(\ref{eq:euler_equation}) in each cell of interest by projecting the state vector~(\ref{eq:state_vector_definition}) onto a set of orthogonal basis functions for each cell. The resulting solution representation is allowed to be discontinuous across element boundaries, i.e.~each cell has its own projection that is independent of that in neighbouring cells. The time evolution of the solution in each cell and the coupling of the solutions across cell boundaries are derived from a weak form of the underlying differential equations. At cell boundaries, this gives rise to numerical flux functions 
that can be computed with the help of Riemann solvers, similarly to how this is done in finite volume discretizations with Godunov's method.

\subsection{Basis expansion}

To be more explicit, we express the state vector $\boldsymbol{u}^K(\boldsymbol{x}, t )$ in each cell $K$  as a linear combination of time-independent, differentiable basis functions $\phi_l^K(\boldsymbol{x})$, \begin{equation}
    \label{eq:state_vector_expansion}
    \boldsymbol{u}^{K}(\boldsymbol{x}, t) = \sum_{l=1}^{N} \boldsymbol{w}_{l}^{K}(t)\, \phi_{l}^{K}(\boldsymbol{x}),
    \end{equation}
where the $\boldsymbol{w}_l^K(t)$ are $N$ time dependent weights. Since the expansion is carried out for each component of our state vector separately, the weights $\boldsymbol{w}_l^K$ are really vector-valued quantities with 5 different values in 3D for each basis function $l$. Each of these components is a single scalar function with support in the cell $K$.

We decompose our simulation domain into a set of non-overlapping cells of equal size, and we pick tensor-products of Legendre polynomials as basis, so that  each cell has a smooth polynomial solution within it. The solution may in general jump across the cell boundaries, and a special treatment is needed for the diffusion equation in this case due to its second spatial derivatives (and likewise for the Navier Stokes equations), which we will briefly specify below. In any case, at a given time the global numerical solution is fully determined by the set of all weights. 

In the following, we only consider Cartesian cells of uniform size and a fixed number of basis functions per cell. It is possible to generalise the DG approach to a variety of other cell geometries, to spatially vary the cell size ($h$-refinement), and to modify the expansion order applied to individual cell's (so-called $p$-refinement). For more details on DG implementations that realize adaptive mesh refinement, see for example \citet{Schaal2015} and \citet{guillet_high_order_2019}. For DG methods with local $p$-refinement see \citet{Mossier2022-br} and references within. 

\subsection{Time evolution}

To evolve the simulation in time we need to derive a way for evolving the time-dependent weights from one time-step to another. 
Starting with the Euler equations~(\ref{eq:euler_equation}), we multiply it with a test function, e.g.~one of our basis functions $\phi_l$, and integrate over a cell $K$, yielding 
\begin{equation}
\label{eq:euler_integrated_over_cell}
\int_K  \phi_{l}^{K}    \frac{\partial \boldsymbol{u}}{\partial t} {\rm d}\boldsymbol{x} +
\int_K
\phi_{l}^{K} \,\boldsymbol{\nabla} \boldsymbol{F} \, {\rm d}\boldsymbol{x} = 0. 
\end{equation}
Integrating the second term by parts and using Gauss's theorem we can transform the integral over the cell into an integral over volume and its outer surfaces, respectively, yielding the so-called weak formulation of the hyperbolic conservation laws of the Euler equations:
\begin{equation}
\label{eq:weakform}
\int_K  \phi_{l}^{K}    \frac{\partial \boldsymbol{u}}{\partial t} {\rm d}\boldsymbol{x} +
\int_{\partial K}
\phi_{l}^{K} \, \boldsymbol{F} \, {\rm d}\boldsymbol{n}  
-
\int_K
\boldsymbol{\nabla} \phi_{l}^{K} \, \boldsymbol{F} \, {\rm d}\boldsymbol{x} = 0,
\end{equation}
where $|K|$ stands for the volume/area/length of the cell.

Using  the orthonormality of our Legendre basis,
\begin{equation}
    \int_K \phi_{l}^{K}(\boldsymbol{x}) \phi_{m}^{K}(\boldsymbol{x}) {\rm d}\boldsymbol{x} = \delta_{l, k} |K|,
\end{equation}
we can simplify the integrals and obtain a differential equation for the time evolution of the weights:
\begin{equation}
\label{eq:weight_evolution}
|K| \frac{{\rm d} \boldsymbol{w}^K_{l}}{{\rm d} t}
= \int_K
\boldsymbol{\nabla} \phi_{l}^{K} \, \boldsymbol{F} \, {\rm d}\boldsymbol{x}
- \int_{\partial K}
\phi_{l}^{K} \, \boldsymbol{F}^\star(\boldsymbol{u}^+, \boldsymbol{u}^-) \, {\rm d}\boldsymbol{n}.
\end{equation}
Here we also considered that the flux function at the surface of cells is not uniquely defined if the states that meet at cell interfaces are discontinuous. We address this by replacing $\boldsymbol{F}(\boldsymbol{u})$ on cell surfaces with a flux function 
    $\boldsymbol{F}^{\star}(\boldsymbol{u}^+, \boldsymbol{u}^-)$ 
that depends on both states at the interface, where $\boldsymbol{u}^+$  is the outwards facing state relative to $\boldsymbol{n}$ (from the neighbouring cell),  and $\boldsymbol{u}^-$ is the state just inside the cell.  We typically use an approximate Riemann solver for determining $\boldsymbol{F}^{\star}$, but of course an exact Riemann solvers can be used as well. In the remainder of this work, we use the
Riemann HLLC solver by \citet{toroRiemannSolvers} as implemented in the {\small AREPO} code \citep{Springel2010, Weinberger2020}.

What remains to be done to make an evaluation of Eqn.~(\ref{eq:weight_evolution}) practical is to approximate both the volume and surface integrals numerically. For the integrations, we employ Gaussian quadrature that turns the volume and surface integrals into discrete sums. The number of Gauss points needs to be chosen consistently with the selected expansion order k \citep[see][]{Schaal2015} such that the $L_1$-error norm,
\begin{equation}
\label{eq:l1_norm}
L_1 =  \frac{1}{|K|} \int_{K} \left|\, \boldsymbol{u}(\boldsymbol{x})  -   \sum_{l=1}^{N(k)} \boldsymbol{w}_{l}^{K}\, \phi_{l}^{K}(\boldsymbol{x})\, \right| \mathrm{~d} V,
\end{equation}
of the total approximation error declines as $L_1 \propto h^{-(k+1)}$ with spatial resolution $h$. Using expansion order $k$ results in a method with a spatial order $p=k+1$. Similarly, the time integration method of the differential equation for ${\rm d} \boldsymbol{w}^K_{l}/{{\rm d} t}$ needs to be of sufficiently high order to avoid that time integration errors dominate the total error budget. We choose Runge-Kutta schemes of appropriate order to achieve this goal. For full details, in particular for the location of the Gauss points and for the specific enumeration of the basis functions we have chosen, we refer to our earlier paper. There, also other practical aspects, such as the definition of the weights for given initial conditions, are discussed.

\subsection{Diffusion operator across cell boundaries}

To generalize the above approach to treat the full Navier-Stokes equations (hereafter NS), or a general diffusion operator $\nabla \cdot(\varepsilon \nabla \boldsymbol{u})$ that we used in our previous work \citep{subsonic_dg_gpu} to introduce artificial viscosity for shock capturing, we add the corresponding dissipative term as a source term to the basic Euler equation, so that it reads, for example, as \begin{equation}
    \label{eq:navier-stokes-bassi}
    \frac{\partial \boldsymbol{u}}{\partial t}+\nabla \cdot \boldsymbol{F}=\nabla \cdot(\varepsilon \nabla \boldsymbol{u}),
\end{equation}
with $\boldsymbol{u}$ being the state vector (\ref{eq:state_vector_expansion}) and $\boldsymbol{F}$ the flux matrix (\ref{eq:flux_matrix}).

The crucial difference between the normal Euler equations~(\ref{eq:euler_equation}) and this dissipative form is the introduction of  a second derivative on the right-hand side, which modifies the character of the problem from being purely hyperbolic to an elliptic type, while retaining manifest conversation of mass, momentum and energy. This second derivative can not be readily accommodated in our weight update equation obtained thus far. Recall, the reason we applied integration by parts and the Gauss' theorem going from Eq.~(\ref{eq:euler_integrated_over_cell}) to Eq.~(\ref{eq:weakform}) was to eliminate the spatial derivative of the fluxes. If we apply the same approach to $\nabla \cdot(\varepsilon \nabla \boldsymbol{u})$ we are still left with one $\nabla$-operator acting on the fluid state.

Our method for addressing this effectively works by constructing a new continuous solution of $\boldsymbol{u}$ across all pairs of adjacent cells. To this end we create a ``virtual'' cell that overlaps partially or in full with the two constituent cells. By evaluating each cell's weights and projecting them onto the common overlapping basis we obtain the basis of the virtual cell. Note that this projection is a sparse matrix operation in which the new coefficients are a sum of the old expansion coefficients, making the estimation of second derivatives at cell interfaces reasonably efficient.

\subsection{Parallelisation on GPUs}

Compared to ordinary finite volume schemes, DG approaches require the evaluation of polynomial expansions at a variety of Gauss points, and the cell evolution is described not only by cell averages but instead by multi-valued expansion vectors for each fluid variable. Calculating the time evolution of these high-order weights increases the computational work needed per cell. At the same time, the coupling to neighbouring cells at arbitrary order only ever involves surface states. In contrast to finite volume codes, where ever deeper stencils are needed for increasingly higher order reconstructions. As such the algorithm therefore features a comparatively high computational intensity with only modest communication needs in comparison to high-order finite volume approaches.  These characteristics  are in principle favourable for reaching a high fraction of the theoretical peak performance on modern computing hardware which operates in a Single Instruction Multiple Data (SIMD) mode. And since much of the work on different cells can be done fully in parallel, it is attractive to consider GPUs as computational engines for DG methods.

We have therefore developed our DG implementation from the ground up to use GPUs. Otherwise, CPUs can also be used. Parallelisation over multiple GPUs is achieved through the message passing interface (MPI), i.e.~clusters of compute nodes each equipped with one or several GPUs can be employed. In principle, our code architecture also allows a mixed operation of CPUs and GPUs, although this is typically not a preferable strategy in practice as their relative speeds will in general not be well matched, and our work-load decomposition between the two is static and fixed at start-up. Full technical details of our code are described in Section~9 of our previous paper \citep{subsonic_dg_gpu}. In the present work we focus primarily on the algorithmic efficiency of DG for problems involving many shocks and not on absolute code speed. The latter is of course also quite sensitive to implementation details and the employed computing hardware.

\section{Method improvements for supersonic hydrodynamics}
\label{Sec:method_improvements_for_supersonic}
In this section, we introduce two changes which allow the DG method to handle highly supersonic, ${\cal M}\sim 12.8$, flows. First we employ the classic von
Neumann-Richtmyer viscosity to prevent the growth of spurious oscillations at shocks. This version of artificial viscosity works by invoking artificial pressure at locations with large negative velocity divergence (i.e. rapid compression), so that the amount of work needed to compress a parcel of gas is increased. As the artificial pressure is removed when the gas expands, this provides irreversible dissipation, thereby helping to stabilize the shock capturing.

The second improvement is a more conservative extrapolation of velocity values to cell interfaces. In standard DG, the velocity field inside cells is defined as the ratio of two separately evolved polynomials representing momentum and mass density, respectively. This ratio is well-behaved at interior Gauss points but is prone to produce abnormally high values when evaluated at the extrapolation points on cell interfaces.
\subsection{Viscous shock capturing}
\label{Sec:Viscosity}

One important conceptual feature of DG is that there is no source of viscosity in the sub-cell evolution, because DG is designed to evolve a
{\it smooth}, differentiable field of the conservative variables under the {\it inviscid} Euler equations as accurately as possible. By construction, there is no source of entropy in this evolution. It follows that a true physical discontinuity, in the form of a shock wave in which the inviscid assumption breaks down, cannot be represented correctly -- because this would require that entropy is produced by {\it irreversibly} converting some of the kinetic energy to heat. 

In our previous study we have addressed this by introducing an explicit viscosity field that was treated with a special high-order DG solver for a diffusive source term added to the Euler equations (i.e.~turning them effectively into a generalized form of the Navier-Stokes equations). This artificial viscosity field could then be used  for the purpose of shock capturing, besides optionally adding physical viscosity and/or heat diffusion. To steer the strength of the artificial viscosity, we had introduced both a simple shock sensor based on the rate of local compression and a `wiggle sensor' that was meant to detect rapid, spurious oscillations in the flow. Each of them could ramp up the local viscosity, while without such a sensor trigger the strength of the artificial viscosity was made to decay again to zero on a short timescale. 

We could demonstrate that this approach allowed a capturing of shock waves at sub-cell resolution. Still, this scheme is quite complicated and technically involved, as the treatment of the viscous source function introduces additional computational cost as well as memory overhead. Another disadvantage is that some of the viscosity was effectively added as a type of post hoc damage control, namely only when the solution already exhibited a strongly oscillatory character. The simulation thus first needed to develop a problematic local character before this is ``healed'' again by supplying needed dissipation, while it would evidently be better to prevent the occurrence of local problems in the first place.

We have therefore reconsidered the parametrisation of our artificial viscosity. One should perhaps first comment that the word ``artificial'' is really a bit of a misnomer in this context. While we stick to using this term for consistency with the literature, a better name would arguably be ``required viscosity'', because having no dissipation in a DG-cell that features a shock is physically plainly wrong. Adding the viscosity that needs to be there is hence in principal ``natural'' not artificial.

In any case, we here resort to a version of the well-known von Neumann-Richtmyer viscosity first described in \citet{VonNeumann1950}, which has been exploited successfully in the field for decades \citep{Wilkins:1980aa}, and incidentally has also motivated the parametrisation of artificial viscosity commonly employed in smoothed particle hydrodynamics \citep{Monaghan1983}. The von Neumann-Richtmyer viscosity is based on the idea to introduce a viscous pressure $\Pi$ in rapidly compressing parts of the flow (indicating a region undergoing a shock), and to add it to the ordinary thermal pressure, so that the sum of the two pressures enters in the usual place in the momentum and energy equations. The effect of this will be that the compression is slowed, with kinetic energy being converted to internal energy in an energy-conserving fashion. However, since the excess pressure $\Pi$ is only added during the compression phase, the produced heat energy does not give rise to the same pressure when the gas can expand again, thus the thermal energy cannot be converted back to kinetic energy in full, unlike for ordinary adiabatic compression and expansion. Such an irreversible conversion of kinetic energy to heat is exactly what happens at a shock, and it is a process that is associated with entropy production.

More explicitly, if we label the entropy per unit mass of the gas through an entropic function, $A = p / \rho^\gamma$, then the Euler equations show that the volume density $\rho A$ of the entropic function is a conserved quantity outside of shocks \citep[e.g.][]{Springel2002}, governed by the additional conservation law
\begin{equation}
\frac{\partial}{\partial t} \left( \rho A\right)  + \vec{\nabla} \cdot\left( \rho A \boldsymbol{v}\right) =0. 
\end{equation}
Adding a viscous pressure in the Euler equations as described above gives rise to
\begin{equation}
\frac{{\rm d}A }{{\rm d} t}  =  - \frac{1}{2}\frac{\gamma-1}{\rho^\gamma}\Pi\, \nabla \cdot \boldsymbol{v}, 
\end{equation}
where ${\rm d}/{{\rm d} t}$ is the convective derivative. Hence, a judiciously chosen $\Pi$ can inject the required entropy.

The basic parametrisation of the von Neumann-Richtmyer viscosity we adopt is the classic form
\begin{equation}
\Pi = \alpha_{\rm visc} \, \rho \left(\frac{h}{p}\right)^2 |\nabla\cdot \boldsymbol{v}|^2,
\end{equation}
for $\nabla\cdot \boldsymbol{v} < 0$, otherwise $\Pi$ is zero. Here $h/p$ gives the expected spatial resolution of our DG scheme of order $p$ (with $h$ being the cell size). The parameter $\alpha_{\rm visc}$ is dimensionless and roughly determines over how many resolution elements a shock is resolved. Typical values should be in the range $\alpha_{\rm visc} \simeq 1.0-3.0$. Note that the precise value will not be important for determining the properties of the post-shock flow, as the total dissipation occurring at a shock is prescribed by the conservation laws, i.e.~the effective shock profile auto-adjusts such that that the correct total dissipation occurs. However, the sharpness of the shock and the degree to which there may be residual post-shock-oscillations still depend on $\alpha_{\rm visc}$ and the functional form adopted for $\Pi$.

The quadratic dependence on $\nabla\cdot \boldsymbol{v}$ proves effective in selectively adding viscosity in shocks, while introducing only negligible viscosity in other places of the flow. However, the above parametrisation can still leave some postshock oscillations downstream of a shock, essentially because the viscosity shuts off too rapidly after passing through the strongest rate of compression. To mitigate this, one can augment the viscosity with a small additional bulk viscosity contribution, of the form
\begin{equation}
\Pi = \beta_{\rm visc\,} \rho c_s \left(\frac{h}{p}\right) |\nabla\cdot \boldsymbol{v}|,
\end{equation}
where $c_s$ is the sound speed. As the latter increases in a shock, this preferentially affects the shock region past the maximum compression rate, and thus helps to damp out postshock oscillations. This viscosity parametrisation is however less specific than that with a quadratic dependence on the velocity divergence, and hence can lead to an unwanted damping of flow features such as sounds waves when used with a non-negligible value of $\beta_{\rm visc}$. We thus typically either set $\beta_{\rm visc}=0$, or choose a value around $\beta_{\rm visc}\simeq 0.1\, \alpha_{\rm visc}$. 

In most practical applications we have found that $\alpha_{\rm visc} \sim 2.0$ and $\beta_{\rm visc} \sim 0.2$ provide a good compromise between stability, narrowness of shocks, and the damping of postshock oscillations, largely independent of flow type and DG-order $p=k+1$. To protect against the possibility that the viscous force applied in one timestep could become so large that it would ${\it reverse}$ the compressive motion, we limit $\Pi$ against a maximum value of
\begin{equation}
\Pi_{\rm max}  = \frac{1}{2} \rho \left(\frac{h}{p}\right)^2 \frac{|\nabla\cdot \boldsymbol{v}|}{\Delta t}   ,
\end{equation}
where $\Delta t$ is the prescribed timestep at the beginning of the step. A similar type of limiter is used in the SPH code {\small GADGET} \citep{Springel2001}.

In practical terms, we simply add $\Pi$  to the pressure computed for the fluid state at all {\it internal} Gauss-points used in the volume integration over the flux function, on the grounds that here the inviscid Euler equations need to be augmented with dissipative terms to introduce entropy production where necessary. $\Pi$ is calculated using the quadrature point specific $\rho$ and $|\nabla \cdot \boldsymbol{v}|$. In the surface integrals, we do not introduce any artificial viscosity. This is because the Riemann solver computes a wave solution that injects entropy into the downstream cell when appropriate, or in other words, here the inviscid Euler equations are implicitly already supplemented with a means to irreversibly convert kinetic energy to heat. 

Recall for comparison that in finite volume methods there are two ways to produce entropy. One is through the Riemann problems solved at cell interfaces, and this is present in equivalent form in our DG approach as just mentioned. The other is through the implicit averaging step that is done at the end of every timestep, where only the average state of cells is retained (to be followed by a reconstruction step from scratch the the beginning of the next step). This averaging step also produces entropy in general, for example when it mixes gas phases of different temperature that have streamed into a cell. The Discontinuous Galerkin approach misses this source of entropy (likewise this is absent in  SPH). In many situations this is advantageous, for example in pure advection, while for shocks it is an  \mbox{impediment -- here DG needs} to be augmented with a suitable channel to entropy production, and this is exactly what we achieve with the artificial viscosity.

In order to be able to directly compare our DG implementation with artificial viscosity shock-capturing to a classic second-order accurate finite volume (FV) scheme, we have added such a scheme to our code as well. The FV approach can in essence be viewed as a DG-scheme of order $p=1$, i.e.~where only cell-averages of the conserved variables are stored, but which is augmented with a reconstruction step that computes linear slopes of the fluid variables for each cell (through piece-wise linear reconstruction), raising it again to the description of the fluid as done by a $p=2$ DG-scheme.  Then these slopes are used to compute the interface states left and right of all cell interfaces, which are in turn fed to the Riemann solver to compute the fluxes between cells. Unlike in a DG scheme of order $p=2$, the slopes are not evolved in time, but rather discarded after every step and then re-estimated. The FV scheme therefore does not need to  compute fluxes inside cells, unlike the corresponding DG scheme. 

For carrying out the piece-wise linear reconstruction in our FV scheme, we estimate the slopes for the primitive variables in each spatial direction, preventing over- and undershoots with a monotonised central slope limiter. This limiter still has the so-called total variation diminishing (TVD) property, but it is substantially less diffusive than, for example, the minmod limiter. Note, however, that the scheme is not guaranteed to be positivity preserving, so that in simulations with extreme density variations (such as in supersonic turbulence) we have introduced an additional slope-liming criterion based on a troubled cell indicator in order to be able to robustly run simulations in all situations. In particular, if a cell ends up with negative density in a timestep, such a cell is flagged as a  `troubled cell' for this step, meaning that its slope estimate is set to zero, and the corresponding timestep calculation is simply repeated. Because for flat slopes positivity can be guaranteed for reasonable timesteps, this then allows the simulation to proceed.

For our general DG implementation, we  employ a similar approach to guarantee positivity and code stability in case challenging local flow situations should arise. We here verify positivity at all Gauss points also in all intermediate steps of the Runge-Kutta time integration. If negative density or pressure values occurs, we apply a positivity limiter that in the first instance tries to scale all high-order weights such that the negative values can be avoided. The computation of the timestep is then repeated. If even a flat expansion inside a cell is not able to rectify the situation, we reduce the timestep size and try again. Before closing this section, we consider two illustrative tests of the new artificial viscosity treatment in DG, starting with the Shu-Osher shock tube.

\subsubsection{Shu-Osher shock tube}
\begin{figure*}
	\includegraphics[width=18cm]{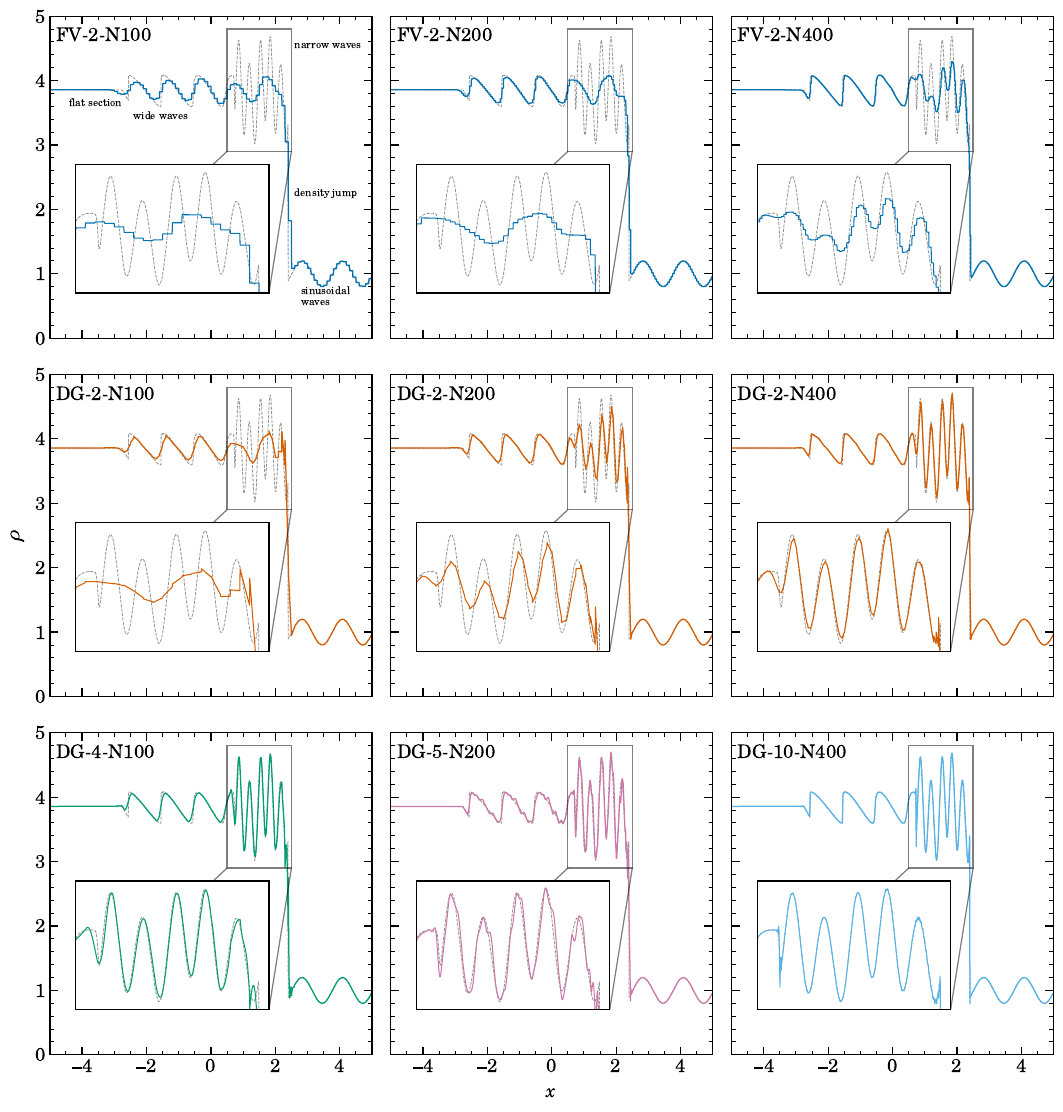}\vspace*{-0.2cm}
    \caption{Shu-Osher shock interaction test problem at time $t=1.8$, for different resolutions and numerical schemes. The initial conditions contain a Mach number ${\cal M}=3$ shock wave that is incident on a sinusoidal density perturbation. The top row shows the problem when simulated at different resolutions (as labelled, where the number following the method name specifies the method's spatial convergence order and the number following `N' is the number of cells over a domain length of 10 units) with a conventional second-order finite volume (FV) method with piece-wise linear reconstruction. Even with 400 cells, the short-wavelength wiggles (see the enlarged insets) in the solution (dotted line) are only poorly resolved. In the middle row, we show equivalent DG computations at order $p=2$, i.e.~also with a linear expansion inside cells. The results especially for the 200 and 400 cell resolutions are drastically improved. In the bottom row, we extend the results to higher order DG schemes, up to a tenth-order accurate scheme ($p=10$), demonstrating that our implementation can robustly treat strong shocks at high order thanks to our new artificial viscosity scheme. }
    \label{fig:shu_osher}
\end{figure*}

In Figure~\ref{fig:shu_osher} the well-known Shu-Osher shock tube problem \citep[][their test problem 8]{Shu1989}. This describes the interaction of a strong incoming Mach ${\cal M} = 3$ shock wave with an adiabatic standing wave in density. The result is a complicated oscillatory pattern in the downstream region of the shock, which is challenging for numerical schemes to resolve accurately. The initial conditions are given by, for $z<-4$, as $\rho=3.857143$, $v_x= 2.629369$, and $P = 10.33333$, and for $x\ge -4$ as  $\rho =1 + 0.2 \sin(5 x)$, $v_x=0$ and $P=1$, with and adiabatic index $\gamma  = 1.4$.

The solution domain at $t=1.8$ has five zones, from left to right they are the flat initial section, followed by wide waves, followed by very sharp narrow waves, then a sharp density jump and finishing with sinusoidal oscillations. 
For $N_\text{cell} = 100$ the finite volume scheme with linear reconstruction is able to reliably resolve zone one and five. The same order DG method shown in the middle row performs slightly better in the wide waves section and much better resolving the last smooth waves section. Going to $p=4$, which results in a 4-th order method shown in the bottom row, the blue line only has slight deviations from the analytic solution in zone two where it fails to resolve the sharp edges of the zig-zag waves. Moving on to $N_\text{cell} = 200$, the FV method resolves the wide zig-zag waves better, while the sharp narrow waves do not improve significantly. On the other hand, the same zone improves significantly with $p=2$ in DG, almost fully resolving the narrow waves. At $N_\text{cell} = 400$, FV can basically fully resolved the zig-zag waves, akin to DG. But it still fails to resolve zone three with its sharp narrow waves. The same order DG method resolves these waves much better while also accounting for the very sharp upper bump between the waves and the shock, albeit with some ringing. In the bottom row we show the performance of high order DG methods in this problem to showcase their stability. It is interesting to see the ringing at $N_\text{cell} = 200$ with $p=6$ at the zig-zag waves and slightly at the narrow waves, while this ringing is missing at $N_\text{cell} = 400$ with order $p=10$. This very high order method does not suffer from any ringing in this problem, showcasing the robustness of our DG  also in the presence of strong shocks at high order, made possible by the use of the artificial viscosity.

\subsubsection{Implostion test}
As second problem we consider the implosion test of \citet{Liska2003}, which consists of a square-shaped 2D domain of extension $[0, 0.3]^2$ with reflective boundary conditions in which the region $x+y <0.15$ has  initially  density $\rho=0.125$ and pressure $P=0.14$, while all the other gas has $\rho =1$ and $P=1.0$. The gas is at rest in the beginning and has an adiabatic index of $\gamma = 1.4$. When the system evolves in time, the region of strongly reduced density and pressure in the lower left corner produces a shock towards the origin which undergoes a double reflection at the domain walls. The interaction of the shocks at the corner and the diagonal produces a jet of dense gas along the diagonal direction. In addition, further shocks bounce off at the opposite sides of the domain, and the  Richtmyer-Meshkov instability produces intricate flow features as shocks cross contact discontinuities in the problem.

\begin{figure*}
\includegraphics[width=180mm]{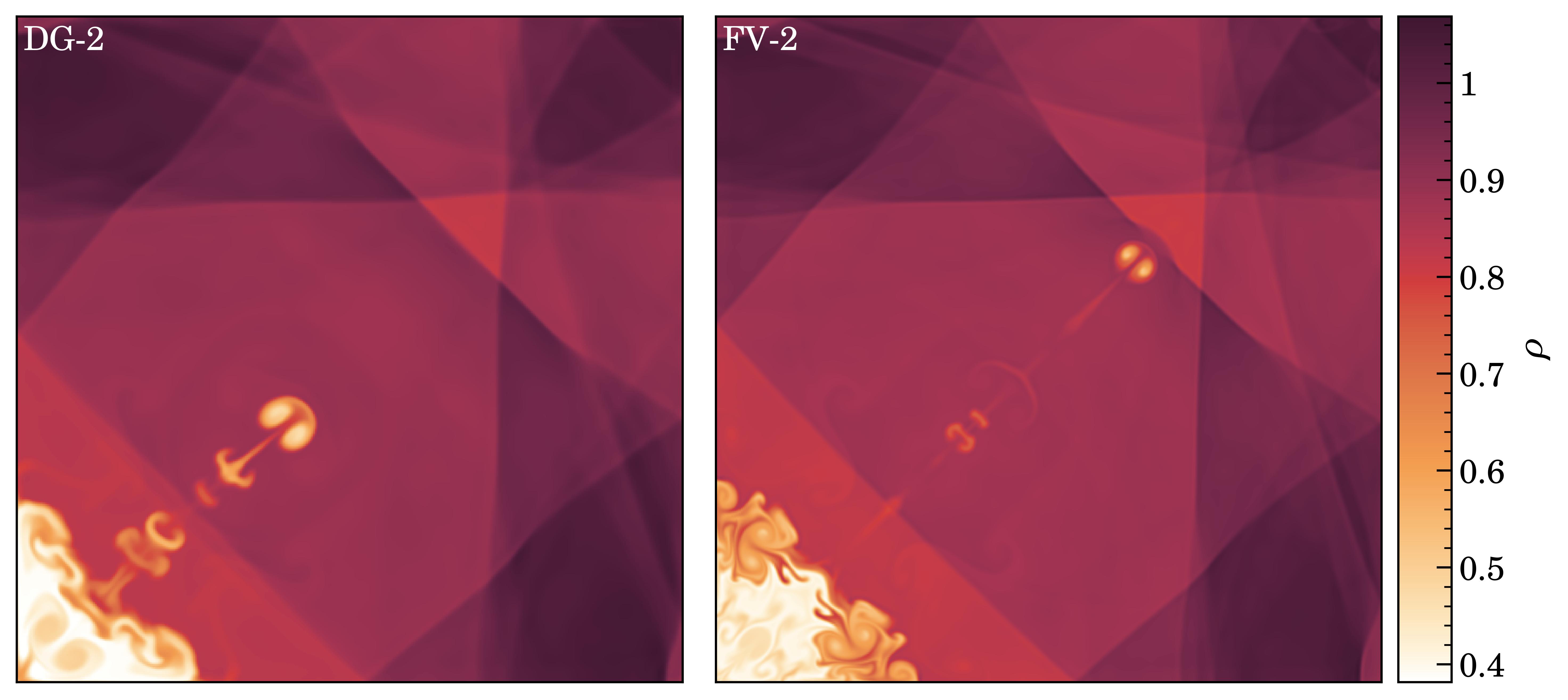}
    \caption{Density field of the \citet{Liska2003} implosion test at time $t=2.5$, simulated with $400\times 400$ cells either with DG at order $p=2$ (right panel), or with a finite volume scheme (left panel). Both methods describe the fluid with linear functions inside cells. The initial conditions contain a region of strongly reduced density and pressure in the lower left corner. This launches a shock towards the origin which reflects at the reflecting boundaries of the domain. The interaction of the shocks at the corner and the diagonal produces a jet of dense gas along the diagonal direction. The test is very sensitive to numerical diffusion, which tends to limit the length of the diagonal jet. As our results demonstrate, our DG scheme is not only capable of capturing the strong shock interactions while accurately maintaining the symmetry of the system, it  also shows clearly less numerical diffusion than the equivalent finite volume scheme.}
    \label{fig:LiskaWendroffImplosion}
\end{figure*}

In Figure~\ref{fig:LiskaWendroffImplosion} we show the state at time $t=2.5$ for a DG simulation with $400\times 400$ cells at order $p=2$ (i.e.~with a linear run of the fluid variables inside cells), and we compare to an equivalent finite volume simulation with the same number of cells. This test is very sensitive to numerical diffusion, which tends to limit the length of the diagonal jet. The comparison highlights that the DG scheme is able to accurately capture the strong shocks in the system based on the artificial viscosity treatment, and it does so with noticeably less numerical diffusivity than the finite volume scheme. In fact, our second-order DG result appears close to or even better than the third-order finite volume result based on piece-wise parabolic reconstruction reported by \citet{Stone2008} for the {\small ATHENA} code.  We have also carried out this test with higher order DG schemes and higher cell resolutions (not shown), which reveal still finer detail in the fluid evolution, as expected.

\subsection{Primitive variables at cell interfaces}
\label{Sec:PrimProjection}

In simulations of supersonic turbulence, extreme density contrasts and networks of strong, interacting and overlapping shocks are encountered that put any numerical scheme to a stress-test in terms of robustness. We already mentioned that this is even the case for simple finite volume schemes that use piece-wise linear reconstructions, but these problems become even more acute in high-order approaches such as our DG scheme. Only extremely diffusive, first order schemes are free of such troubles.

One particular issue we noticed in supersonic turbulence calculations with DG is that our default approach to compute the primitive variables at the cell interfaces can sometimes produce extreme velocity values that are basically unphysical. After being processed by the Riemann solver, the resulting inaccurate fluxes then pollute the solutions inside the cells. The problem originates in our definition of the velocity field inside cells as ratio of the polynomial describing the momentum density and the polynomial describing the density, which both are evolved separately. The ratio of two polynomials is a rational function that can be well outside the space of our underlying polynomial basis functions. While the values obtained at the interior Gauss points should be reasonably well behaved, because these velocities enter the internal flux computation, the {\it extrapolated} values at cell interfaces are much less well constrained. And indeed, in cells that show large excursions of density and/or momentum density from the mean (perhaps even in opposite directions), the velocities one obtains at cell interfaces by dividing the two polynomial expansions can become quite extreme, especially if the density itself approaches very small values.

\begin{figure*}
	\includegraphics[width=18cm]{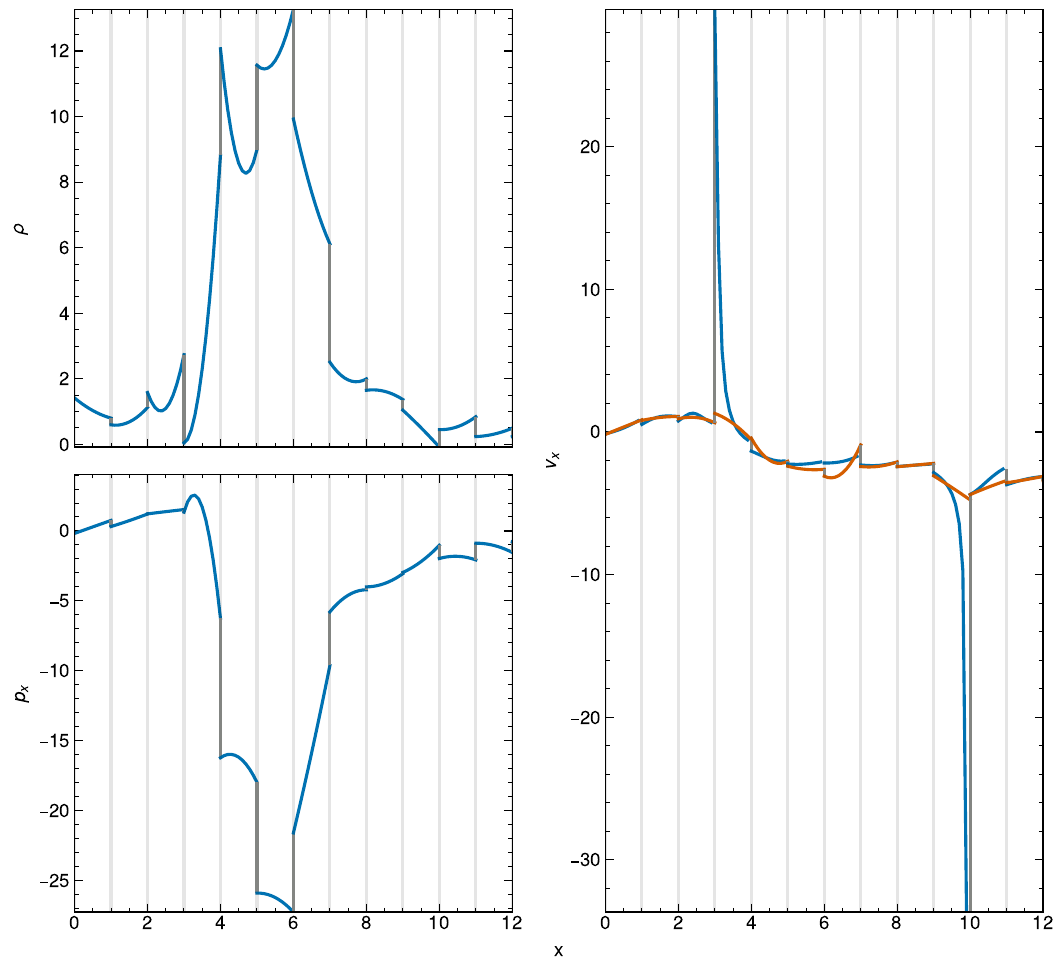}
    \caption{Illustration of the occurrence of problematic, extrapolated primitive variable values at cell boundaries (dark grey) when derived naively from the conservative variables. All panels show a skewer through a 3D, driven-turbulence simulation of high Mach number with vertical lines delineating different cells. The upper left panel shows density, the lower left panel shows the momentum $p_x$ along the $x$-direction. The right panel displays the velocity (blue lines) calculated by taking the ratio of the left panels. This is compared to the velocity calculated with our new method (described in Sec.~\ref{Sec:PrimProjection}), shown in orange. The latter approach projects the velocity itself on the polynomial basis, based on the values attained at the internal Gauss points within a cell.}
    \label{fig:vel_projection}
\end{figure*}

This is illustrated in Figure~\ref{fig:vel_projection} along a one-dimensional skewer through a low-resolution DG simulation ($32^3$ cells with expansion order $p=3$) of Mach number ${\mathcal{M}}\simeq 3$ turbulence. The two panels on the left show the mass density and the $x$-component of the momentum density, respectively, while the right panel shows their ratio (black), i.e.~the inferred $v_x$-velocity. The  cell boundaries are indicated with vertical dotted lines. 

It is clearly seen, and expected, that the polynomial solutions inside cells can in general give rise to discontinuous jumps of the conserved and primitive variables at cell boundaries. These jumps are no problem for the DG scheme, and they preferentially tend to occur when there are shocks and contact discontinuities. However, what is nevertheless a problem are the extreme velocity values that can result at cell interfaces when the momentum and density values are divided by each other, as seen for example in the two cells between $x=6$ and $x=7$, and $x=11$ and $x=12$, respectively. 

We have solved this issue by defining reconstructed primitive variable fields in each cell, simply by computing a polynomial expansion of the primitive variables themselves based on the values they assume at the internal Gauss points of the cells. The procedure for obtaining the corresponding coefficients is akin to how one would project initial conditions onto the polynomial expansion by exploiting the completeness of the basis. The projection of an arbitrary field $\boldsymbol{f}(\boldsymbol{x})$ onto the basis functions $\phi_l^{K}$ of a cell can be done by 
\begin{equation}
\label{eq:weights_equation}
\boldsymbol{w}_{l}^{K} =  \frac{1}{|K|} \int_{K} \boldsymbol{f}\, \phi_{l}^{K} \mathrm{~d} V,
\end{equation}
and the volume integration can be approximated by Gaussian quadrature. We can now set for $\boldsymbol{f}$ the vector of primitive variable fields expressed through the polynomial expansion of the conserved variables (i.e.~for the velocity this will be a rational function obtained as the ratio of momentum density and mass density), and use our standard order for approximating the volume integral with Gaussian quadrature, so that ultimately only the values of the conservative variables at the Gauss points enter. This will then mean that we, for example, obtain an approximation for the velocity field by a polynomial of the same order as used for the conservative variables, and this polynomial goes in principle\footnote{Since in 2D and 3D we discard in the tensor product of Legendre polynomials cross terms of higher order than imposed in 1D, this is not exactly true, and the polynomial projection in essence entails some small level of smoothing.} through the velocity values obtained at the Gauss points, whereas the extrapolated values at the cell interfaces are now bounded and in general better behaved than obtained for the ratio of momentum density and mass density at these same points. Note that for the density field itself, this procedure just returns the same field again, because the density is already a representable polynomial function at the given order. 

When we use these `polynomial extrapolations' for the flux computation at cell interfaces we find that this drastically improves the robustness and the quality of results for our supersonic turbulence and strong shock simulations, whereas for all smooth problems it does not make any tangible difference. Figure~\ref{fig:vel_projection} illustrates this clearly by also including the projected velocity field obtained in this fashion. At those cell interfaces where the ratio of momentum and mass density produced large excursions of the predicted velocities the extrapolations obtained from the projected velocity field are much more reasonable and well behaved. Everywhere else there is no substantial difference.

\begin{figure*}
\begin{center}
    \includegraphics[width=180mm]{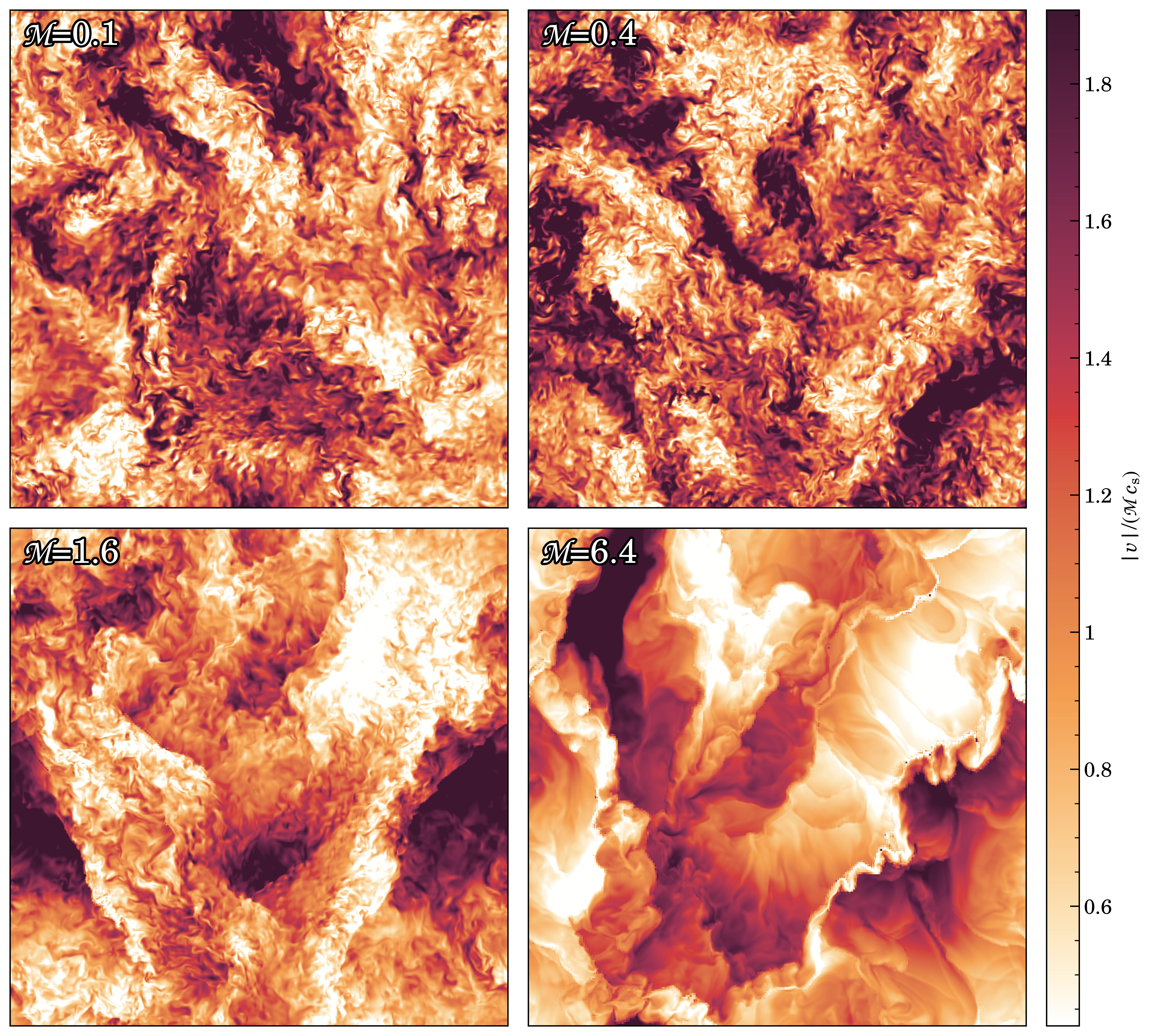}
\end{center}
    \caption{
Slices through the turbulent velocity field of simulations with different Mach number, here ${\cal M}=0.1$, ${\cal M}=0.4$, ${\cal M}=1.6$, and ${\cal M}=6.4$, as labelled. In each case, the color map shows  the velocity amplitude \mbox{$|v| = ({v_x^2 + v_y^2 + v_z^2})^{1/2}$} in units of the corresponding characteristic velocity, here taken as the Mach number times the sound speed. For definiteness, the DG calculations have used  $256^3$ cells and $p=3$, and each panel shows the state after the same number of eddy turn-over times after the start of the simulations. The subsonic simulations show a nearly self-similar structure, as expected for this setup. However, as we transition into the supersonic regime, it is evident that the character of the turbulence qualitatively changes from a box with gas sloshing around to a box defined by a network of overlapping shocks.}
    \label{fig:slice_automatic_vlim}
\end{figure*}

\begin{figure*}
\begin{center}
    \includegraphics[width=170mm]{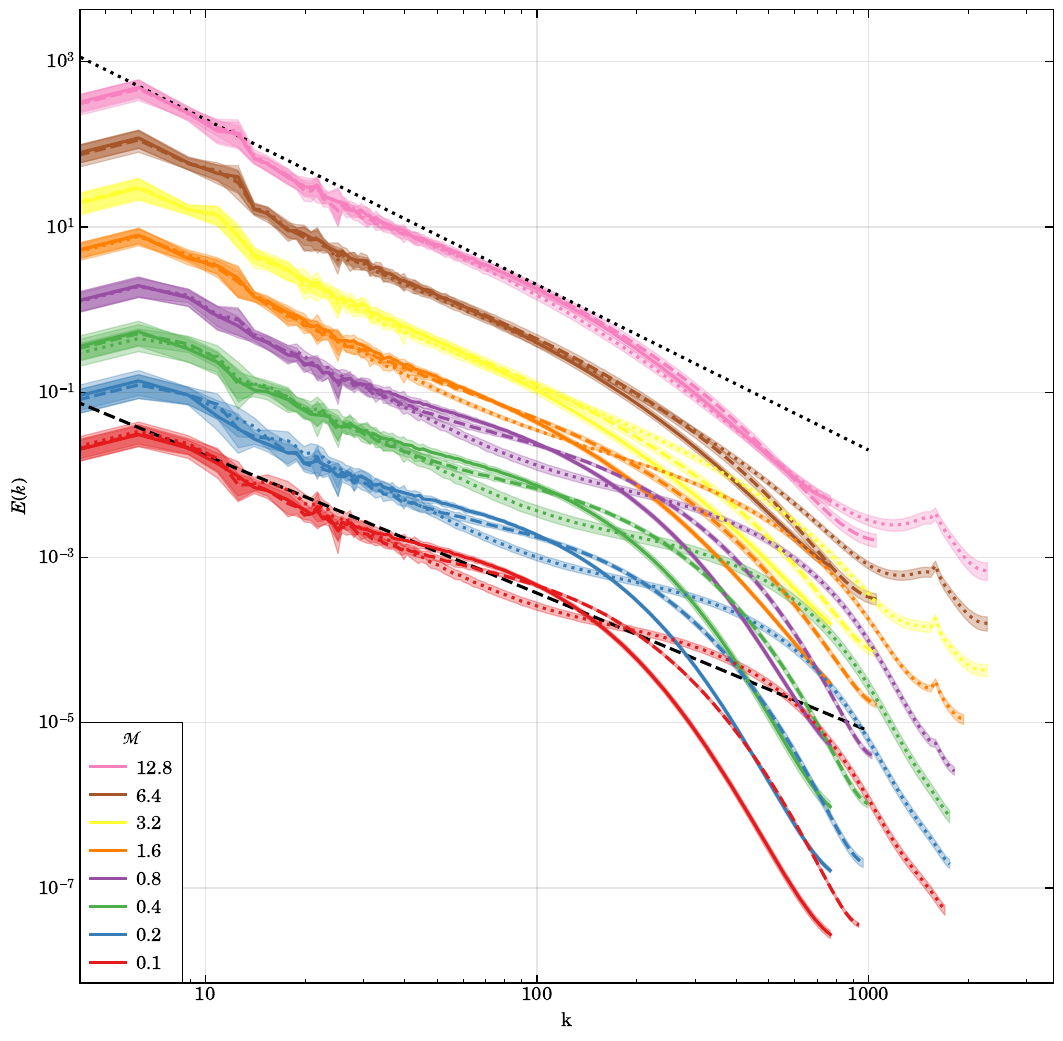}
    \end{center}
    \caption{Velocity power spectra for different turbulent Mach numbers, from the subsonic to the highly supersonic regime, as labelled. The lines show the average of 128 power spectra measurements over 5 eddy turnover times and the shaded regions indicate the $1\sigma$ deviation. For each driving strength, we compare DG simulations with order $p=2$ (dashed) and $p=3$ (dotted) with corresponding finite volume simulation (solid). The black dashed line indicates the Kolmogorov $E(k) \propto k^{-5/3}$ power-law slope, indicative of the subsonic cascade, whereas the dotted black line shows the Burgers $E(k) \propto k^{-2}$ scaling indicative of supersonic turbulence where dissipation is part of the self-similar cascade. The simulations here use only $256^3$ cells and thus have a fairly limited dynamic range that can only resolve a very small part of the turbulent cascade before entering the dissipative regime. Nevertheless, the sequence clearly shows a steeping of the slope towards the supersonic regime, marking the transition from Kolmogorov to Burgers turbulence. Also, the second-order DG runs can resolve the turbulence to higher wave number than the second-order accurate finite volume scheme, reflecting DG's higher accuracy and reduced numerical dissipation. Interestingly, while third-order DG likewise does better than second-order DG in the subsonic regime, this advantage nearly vanishes in the supersonic regime. }
    \label{fig:different_mach_numbers_vel_spectra}
\end{figure*}

\begin{figure*}
\begin{center}
    \includegraphics[width=170mm]{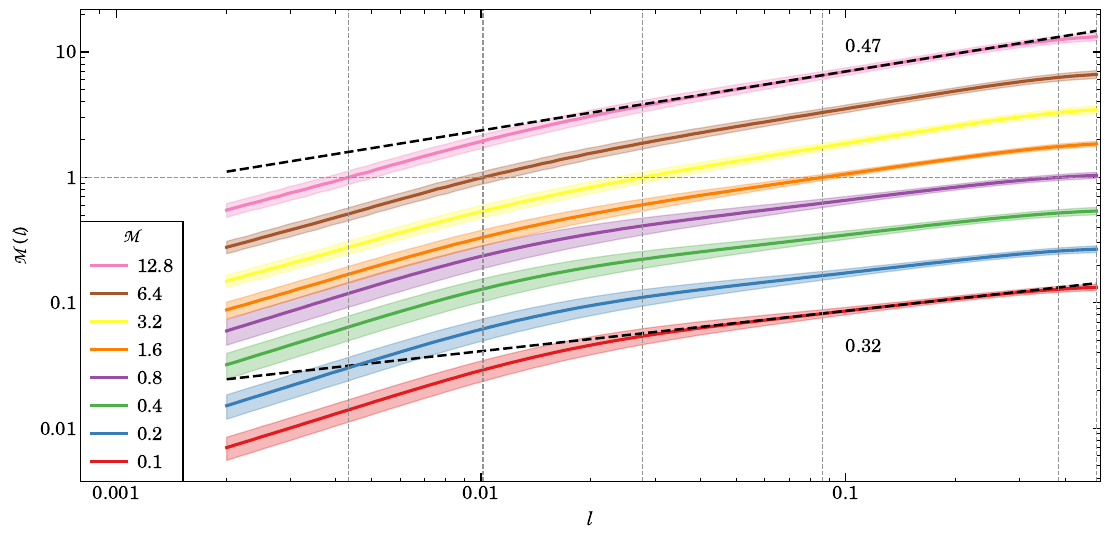}
    \end{center}
    \caption{Velocity structure function for different turbulent Mach numbers, from the subsonic to the highly supersonic regime, as labelled, as a function of spatial scale. The lines show the average of 96 power spectra measurements over 5 eddy turnover times and the shaded regions indicate the $1\sigma$ deviation. For each driving strength, we show DG simulations with order $p=2$. The black dashed lines indicate fits done between $0.1<l<0.25$ for the most subsonic and the most highly supersonic runs. The vertical and horizontal dotted grey lines indicate the super- to subsonic transitions for simulations where it happens. The simulations here use only $256^3$ cells and thus have a fairly limited dynamic range that can only resolve a very small part of the turbulent cascade before entering the dissipative regime. Nevertheless, the sequence clearly shows a steeping of the slope towards the supersonic regime, marking the transition from Kolmogorov to Burgers turbulence. In particular, we measure slopes of 0.32  and 0.47 for our two fits, quite close to the expected scalings of $1/3$ and $1/2$ for subsonic and supersonic turbulence, respectively.}
    \label{fig:different_mach_numbers_vel_structure_function}
\end{figure*}

\begin{figure*}
    \includegraphics[width=180mm]{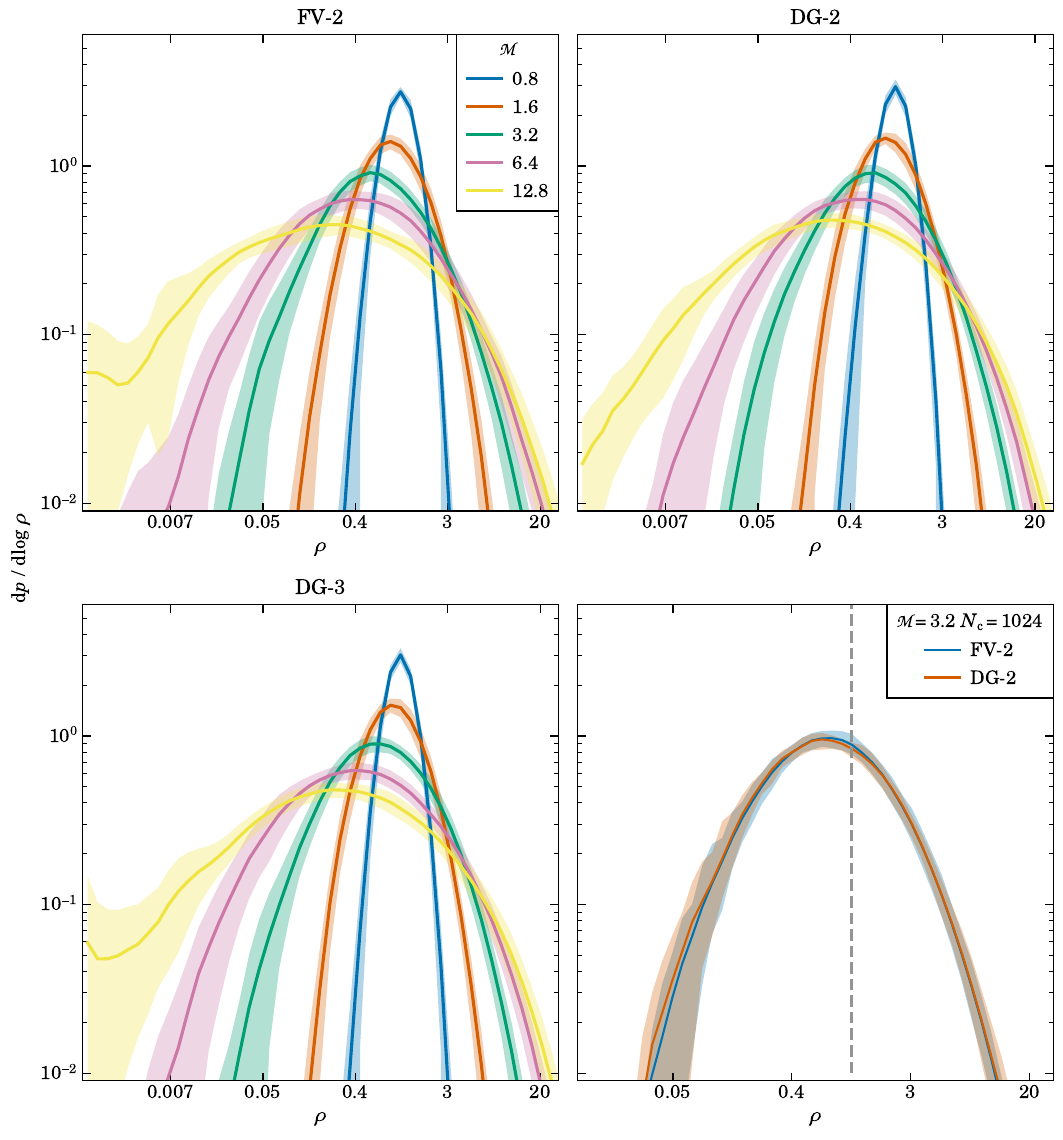}
    \caption{Density probability distribution functions (PDFs) in different turbulence simulations, carried out for a variety of Mach numbers and different numerical schemes. The lines show the average of 96 measurements of three 2D density slices, one per coordinate plane going through the center of the box, over five eddy turnover times, and the shaded regions indicate the $1\sigma$ deviations. In the top two and the bottom left panel, we compare 2$^{\rm nd}$ order finite volume with linear slope reconstruction FV-2, DG at order $p = 2$, and DG at order p=3, as labelled, for a suite of $256^3$ simulations at  Mach numbers from 0.8 to 12.8. All three numerical schemes show a qualitatively very similar behaviour in which the shape of the density PDF transitions from an approximately normal form in  density in the subsonic regime to a log-normal shape in the supersonic regime (note that we use $\log_{10}$ in the PDF's vertical normalization), with a width that grows with Mach number. The bottom right panel compares PDFs at a fixed Mach number of ${\cal M}=3.2$ for higher resolution runs of $1024^3$ cells carried out with FV-2 and DG-2. The ${\cal M}=12.8$ runs for FV-2 and DG-3 show a higher than expected scatter at low densities. This is due to the density floor which we set at $10^{-3}$. This behaviour is not present in the DG-2 run. Here we see that the PDFs are not identical after all, but that the DG scheme is able to resolve a slightly larger density contrast than the corresponding FV scheme.}
    \label{fig:densityPDF}
\end{figure*}

\begin{figure}
\begin{center}
    \includegraphics[width=88mm]{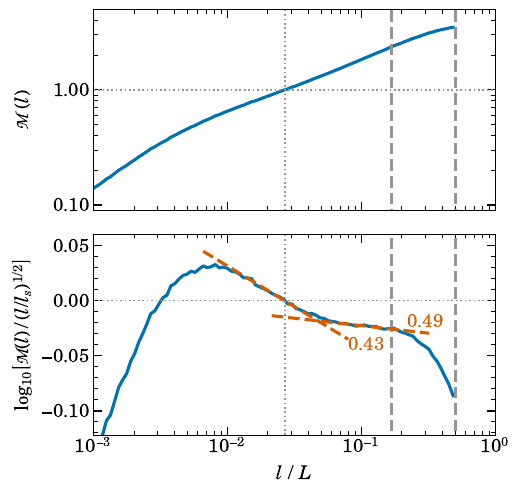}
    \end{center}
    \caption{Velocity structure function (top panel) for a high-resolution DG run with $1024^3$ cells and $p=2$, for driven turbulence with Mach number ${\cal M}=3.2$. The lines show the average of 96 measurements over five eddy turnover times and the shaded regions indicate the $1\sigma$ deviation. We omit the shaded regions in the lower panel for clarity. For pair distances equal to half the box size (right-most dashed vertical line), the structure function starts out at values close to the box-averaged Mach number. From this driving scale, it takes until at least three times smaller scales (marked by the middle dashed vertical line) before a self-similar turbulent cascade develops. The structure function then first drops relatively steeply towards smaller scales, close to the expected $M(l) \propto l^{1/2}$ scaling for Burgers turbulence. Around the sonic point at $l_s$, where ${\cal M}(l_s)=1$, the scaling flattens as the turbulence transitions into the subsonic regime. Here a scaling {$\cal M$}$(l)\propto l^{1/3}$ would be expected if an extended inertial range is present, until a strong steeping sets in when the dissipation regime is entered. The bottom panel shows the velocity structure function in a compensated form, where it is multiplied by the factor $(l/l_s)^{-0.5}$ which brings out subtle shape difference more clearly. Right when the supersonic turbulence cascade sets in, we measure a slope of $0.49$ for ${\cal M}(l)$, close to the expectation. Furthermore, there is a clear break around the sonic scale where the structure function flattens. Our fit in this region returns a slope of $0.43$, somewhat steeper than expected. However, this is not really surprising as the still fairly limited dynamic range of this calculation  and the influence of the bottleneck effect are likely causes for this small difference.
}
\label{fig:compensated_structure_function}
\end{figure}

\section{Driving and measuring turbulence}
\label{sec:supersonic_driving}

In the following, we collect the definitions of some basic quantities to characterize the statistical properties of turbulence and describe our method for driving turbulence. We also detail our measurement techniques for turbulence-related quantities that we examine later on.

\subsection{Basic statistics of supersonic and subsonic turbulence}

The Mach number of turbulence is often defined as
\begin{equation}
 \mathcal{M}  =  \left< \boldsymbol{v}^2 / c_s^2 \right>^{1/2},
\end{equation}
which is a volume-weighted quantity that only depends on the velocity field $\boldsymbol{v}$ in units of the sound-speed $c_s$. It is also possible to define a density-weighted Mach number, given by  $\mathcal{M}_{\rho}  =  \left< \rho \boldsymbol{v}^2/ \left( \overline{\rho} c_s \right) \right>^{1/2} $, which can also be expressed in terms of the total kinetic energy of the flow,
$\mathcal{M}_{\rho} = (2 E_{\rm kin} / M_{\rm tot})^{1/2} / c_s$.
While for subsonic turbulence both measures are equal, $\mathcal{M} \simeq  \mathcal{M}_{\rho}$, for supersonic turbulence there is a small difference, with $\mathcal{M}_{\rho}$ being generally slightly smaller than  $\mathcal{M}$. \citet{Pan:2022aa} cite $\mathcal{M}_{\rho} \simeq \mathcal{M}^{0.96}$ for the relation between the two quantities in the supersonic regime, which matches our own findings very closely.
Note that we consider in this paper only isothermal flows in which $c_s$ is constant, which simplifies the discussion considerably.

The characteristic velocity, $v(l)$, of the turbulent velocity field is scale-dependent, and we define this quantity \citep[following][]{Federrath2021} in terms of the total second-order velocity structure function, as follows
  \begin{equation}
    v(l) =  \left[\frac{1}{2}\left< |\boldsymbol{v} (\boldsymbol{x}) - \boldsymbol{v}(\boldsymbol{x} + \boldsymbol{l})|^2  \right>_{\boldsymbol{x}, l = |\boldsymbol{l}|} \right]^{1/2},
    \label{eqnstrfnc}
\end{equation}
where we average over a large number of random pairs that are separated by a fixed distance $l = |\boldsymbol{l}|$. Based on this quantity, we can also define a scale-dependent Mach number, $\mathcal{M}(l) = v(l) / c_s$.

On the largest scales, we expect $v(l)$ to approach the root-mean-square velocity dispersion $v_{\rm rms}$. To show this, we start with the definition of the second-order structure function

\begin{equation}
    \left\langle\left(\boldsymbol{v}(\boldsymbol{x})-\boldsymbol{v}(\boldsymbol{x}+\boldsymbol{\ell}\right)^2\right\rangle=2 \left( \left\langle \boldsymbol{v}(\boldsymbol{x})^2\right\rangle -  \left\langle \boldsymbol{v}(\boldsymbol{x}) \,\boldsymbol{v}(\boldsymbol{x}+\boldsymbol{\ell})\right\rangle\right).
\end{equation}

On the largest scales where we drive our turbulence, the last term, the autocorrelation, disappears. We therefore simply obtain

\begin{equation}
    \left\langle\left(\boldsymbol{v}(\boldsymbol{x})-\boldsymbol{v}(\boldsymbol{x}+\boldsymbol{\ell})\right)^2\right\rangle_{\ell \rightarrow L/2}= 2 \left\langle \boldsymbol{v}(\boldsymbol{x})^2\right\rangle =2\, v_\textrm{rms}^2 .
\end{equation}

Previous work has demonstrated a scaling of ${v(l) \propto l^\alpha}$ with ${\alpha\simeq {1/2}}$ in the supersonic regime, while this flattens to ${\alpha = 1/3}$ in the subsonic regime. If turbulence is supersonic on the largest length scales, we thus expect the existence of a ``sonic scale'' $l_s$ where the characteristic velocities have fallen to the sound speed, with ${\mathcal{M}(l_s) = 1}$. Based on the velocity scaling in the supersonic regime, we expect this roughly for ${l_s = L_{\rm inj} / \mathcal{M}^2}$, or in terms of wave number, for
\begin{equation}
    k_s \simeq k_{\rm inj} \mathcal{M}^2.
\label{loc_sonic_point}
\end{equation}
This estimate assumes that the supersonic cascade is already well developed right at the injection scale, which is however typically not the case in practice as some range of scales is required before the self-similarity of the cascade is fully established. In any case, unambiguously identifying the sonic point in a turbulence calculation is challenging as it requires to resolve an inertial range {\it both} in the supersonic regime and in the subsonic regime, which demands very high dynamic range. \citet{Federrath2021} have recently accomplished this in a simulation of ground-breaking size, using a grid size of $10048^3$ cells. We shall later try to identify the sonic scale in DG simulations of considerably smaller size. 

Besides characterizing the statistics of the velocity field in real-space through structure functions, it is also common to consider its correlation functions, for example the two-point correlations $\left<\boldsymbol{v}(\boldsymbol{x}+\boldsymbol{l}) \,\boldsymbol{v} (\boldsymbol{x})\right>_{\boldsymbol{x}}$ and its Fourier-transform, the velocity power spectrum. The latter can be defined as 
\begin{equation}
E_v(\boldsymbol{k}) = \left(\frac{2\pi}{L}\right)^3 | \boldsymbol{\hat{v}}(\boldsymbol{k})|^2, 
\end{equation}
where $\boldsymbol{\hat{v}}$ is the Fourier transform of the velocity field. For a statistically isotropic velocity field, it is customary to define the $\boldsymbol{k}$-shell averaged velocity spectrum $E(k)$ through
\begin{equation}
  E(k) = 4\pi k^2 \left< E_v(\boldsymbol{k})\right>.
  \label{eqnvelpower}
\end{equation}
The total velocity dispersion is then given as the integral over $E(k)$. In particular we have
\begin{equation}
\mathcal{M} = \sqrt{ \frac{1}{c_s^2}  \int E(k)\, {\rm d}k} = \sqrt{ \frac{1}{c_s^2}  \left\langle v^2\right\rangle_{\mathcal{V}} }\,.
\end{equation}
In the subsonic regime, we expect the \citet{Kolmogorov1941} scaling $E(k) \propto k ^{-5/3}$ of the velocity power spectrum, while in the supersonic regime this is expected to steepen to \citet{Burgers1948} turbulence with $E(k) \propto k ^{-2}$.

\subsection{Driving isothermal turbulence}

We drive turbulence following the same approach as in our previous work on subsonic turbulence \citep{subsonic_dg_gpu} which in turn follows closely the  procedure described  in many previous works, such as \citet{Schmidt2006,Federrath2008,Federrath2009,Federrath2010,Price2010} and \citet{BauerSpringel2012}.

The acceleration field is constructed in Fourier space between the fundamental mode of the box, $k_{\rm min}=2\pi / L$,  and $k_{\rm max}= 4 \pi / L = 2 k_{\rm min} $, with Fourier mode phases chosen at random from an Ornstein--Uhlenbeck process. As injection scale we can thus define $k_{\rm inj}\simeq k_{\rm max}$, corresponding  to half the box size in real space. The Ornstein--Uhlenbeck process is used because it is temporally homogeneous, meaning its variance and mean remain constant over time. This type of frequent but correlated driving results it a semi-stationary turbulent field which simplifies its sampling. The randomly chosen phases are updated every timestep $\Delta t$, yielding a discrete time evolution update prescription for the Fourier phases $\vect{x}_{t}$, as follows:
\begin{equation}
\label{eq:ornstein_uhlenbeck_process_time_evolution}
\vect{x}_{t}=f\, \vect{x}_{t-\Delta t} + \sigma \sqrt{(1-f^2)}\, \vect{z}_n ,
\end{equation}
where $f$ is a decay factor defined as $f=\exp(-\Delta t/t_c)$, with $t_c$ being the correlation time-scale. $\vect{z}_n$ is a Gaussian random variable and $\sigma$ is the variance.

Through the use of a Helmholtz decomposition, the driving can be made either fully solenoidal, fully compressive, or a combination of the two. To stay consistent with our previous work on subsonic turbulence \citep{subsonic_dg_gpu} we retain the same purely solenoidal driving. In the subsonic regime, compression modes created by compressive driving would result in sound waves propagating through the simulation. Such large-scale sound waves start coupling to smaller scales only when their non-linear steepening starts to dominate. For supersonic turbulence, compressive driving has however a more important influence on the properties of turbulence \citep[e.g][]{Federrath2010,Federrath:2013aa}.

The driving has three free parameters which have to be chosen carefully to quickly establish a quasi-stationary turbulent field that faithfully represents the statistical properties of turbulence at the intended Mach number.
We can define the eddy turn-over timescale on the injection scale as
\begin{equation}
\label{eq:eddy_turnover_time}
    T = \frac{L}{2 c_s \mathcal{M}}.
  \end{equation}
The correlation timescale $t_c$ in the driving prescription is the characteristic lifetime of Fourier modes of the driving field, and thus should ideally be of the order or slightly smaller than the eddy turnover time. Based on this we set the correlation timescale as $t_c \simeq T$. We furthermore set the mode update frequency $\Delta t$ to be 100 times smaller than $t_c$ to assure a smooth transition from one mode to the next. 

The third parameter in Eqn.~(\ref{eq:ornstein_uhlenbeck_process_time_evolution}), $\sigma$, determines the strength of the turbulent driving, and as such the achieved Mach number. To get intuition for this parameter, let us consider the relation between the driving strength and the Mach number in the quasi-stationary end state. We start with the energy injection rate per unit mass, $\epsilon$, which scales as
\begin{equation}
    \epsilon \propto \sigma^2\, \Delta t
 \end{equation}
based on the driving prescription itself. Guided by this expression, we define the strength of the driving through a parameter 
  \begin{equation}
   E_{\rm inj} =  \sigma^2\, t_{c},
 \end{equation}
and express $\sigma$ in terms of $E_{\rm inj}$. Then the achieved energy injection rate scales linearly with the prescribed parameter $E_{\rm inj}$ as
\begin{equation}
    \label{eq:epsilon_propto_einj}
    \epsilon \propto E_{\rm inj} ,
\end{equation}
approximately independently of $t_c$. In the regime of subsonic Kolmogorov turbulence, the driving creates characteristic velocities that are expected to scale with length scale and energy injection rate as
\begin{equation}
    v(l) \propto \left( \epsilon l \right)^{1/3},
\end{equation}
which means that the achieved Mach number in Kolmogorov type turbulence varies as
\begin{equation}
   \mathcal{M} \propto \epsilon^{1/3} \propto E_{\rm inj}^{1/3}.
\end{equation}
For doubling the Mach number, we thus need to triple our driving strength  $E_{\rm inj}$ while $t_c$ and $\Delta t$ should be halved. 

After driving sets in from gas at rest, turbulence tends to become fully developed only for times $t \ge 2 T$. Where $T$ is the large eddy turnover time defined in Eqn.~(\ref{eq:eddy_turnover_time}). We thus analyze turbulence by averaging the results for a large number of outputs between $3\, T < t < 8\, T$ \citep[as, e.g., in][]{Federrath:2013aa}, which gives us enough independent samplings of the box to get robust and converged results with respect to temporal averaging. We note that it also requires some range of scales between the driving scale and the onset of a fully developed self-similar turbulent cascade. One can therefore not expect to immediately obtain proper turbulent scaling right at the scale where the driving ends, but rather needs to go to somewhat smaller scales. For example, \citet{Federrath2021} conservatively estimate that the turbulent supersonic cascade becomes fully developed for $l < L/8$ when the driving is centered at a scale of $L/2$, which is similar to our work.

\begin{figure}
    \includegraphics[width=88mm]{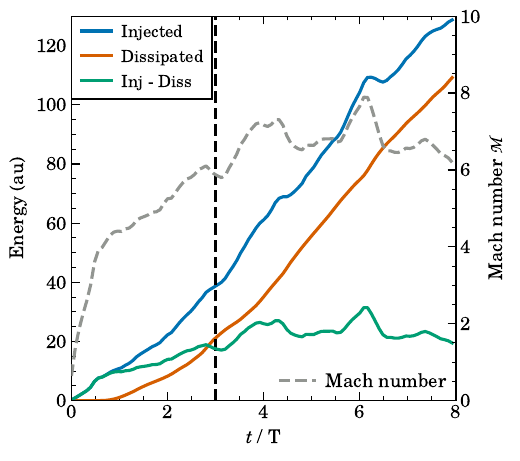}
    \caption{Cumulative injected, dissipated energy and their difference, as well as global volume averaged Mach number, as a function of time in one of our driven turbulence simulations. The vertical dashed line indicates the time at which we start our power spectra measurements. The gas is initially at rest, and put into motion by the driving. Eventually, energy injection is balanced by dissipation in a time-averaged fashion, and the difference between the cumulative injected and dissipated energy is reflected in the kinetic energy as measured by the Mach number.}
    \label{fig:density_pressure_injection_example}
\end{figure}

We note that we typically do not simulate an isothermal gas directly in this work, but rather one with an adiabatic index of $\gamma = 1.0001$. After every timestep, we restore a uniform temperature by extracting (or adding) thermal energy as needed. This allows us to measure the actual energy injection rate by determining the volume integral of the work the external driving field does on the gas, and a dissipation rate by accounting for the energy we need to extract to maintain a uniform temperature. The difference between these time integrated rates is then the instantaneous total turbulent kinetic energy of the gas if it started from rest. In Figure~\ref{fig:density_pressure_injection_example} we show an example of the time evolution of the total Mach number and the cumulative injected and dissipated energies for a turbulence simulation with Mach number ${\cal M}\simeq 6.4$. We see that the cumulative injected energy grows approximately linearly with time, and this evolution is tracked by the dissipated energy, albeit with some time delay. Dissipation only starts to set in after about one eddy turnover time, while it takes until  about $t\simeq 3\,t_{\rm eddy}$ before a quasi-stationary turbulent state has developed where the Mach number does not grow anymore but rather fluctuates around a long-term average value.

\subsection{Measuring structure functions and power spectra}

To measure velocity structure functions, we define a logarithmic set of radial bins between half the box size as maximum distance, and the nominal resolution limit, $L/[pN]$, as minimum distance, where $L$ is the box size, $N$ is the number of cells per dimension, and $k$ the maximum polynomial order of the DG method. Typically we adopt 100 such bins. Upon each output time, we then draw for each bin a fixed number (typically $10^5$) of random positions in the box, plus a set of random directions uniformly distributed over the unit sphere. A second position paired with each point is then determined relative to the first one based on these directions with the distance of the corresponding radial bin, taking periodic wrap-around in the simulation box into account. The velocity values at the two selected coordinates are then evaluated based on the polynomial expansions of the cells the points fall into. The corresponding squared velocity differences are summed for each bin, averaged, and converted to the velocity structure function as defined in Eqn.~(\ref{eqnstrfnc}). To reduce statistical noise in the measurement for a single output and obtain a robust statistical characterization of the quasi-stationary turbulent state, 96 measurements over five eddy turnover times an extended time period are averaged.

For measuring velocity field power spectra, we adopt a Fourier mesh with dimension $N_{\rm grid} = N p$. Velocity values at the regular grid positions are then evaluated based on the polynomial expansions within the corresponding DG cells. We use the MPI-parallel FFTW library to transform the velocity field to Fourier space, separately for each spatial velocity dimension. In each case, the corresponding velocity mode powers are summed up in finely binned $k$-space shells, so that the average mode power of the 3D velocity field can be computed. We typically use a fine set of 2000 logarithmically spaced bins in $k$-space. These fine shells can later be adaptively rebinned in a plotting script to form larger bins, as desired. Here we typically rebin such that bins containing just a few modes are accepted for low-$k$ (otherwise the $k$-bins would get too wide there), while for high-$k$ the bins can be made narrower while still having large mode counts and thus good statistics. Following standard conventions in the field, we present the velocity power spectrum in terms of the quantity $E(k)$ as defined in Eqn.~(\ref{eqnvelpower}).

\section{Turbulence with DG in the supersonic and subsonic regimes}
\label{sec:turbulenceruns}

In this section we want to investigate whether our new DG scheme -- with artificial viscosity shock capturing and an auxiliary projection of the primitive variables to deal with extrapolations to cell boundaries -- is capable of robustly and accurately simulating driven isothermal turbulence well into the supersonic regime.  To this end we first consider a set of simulations where we vary the Mach number systematically but keep otherwise all relevant numerical parameters the same. For definiteness we consider $N^3=256^3$ cells and DG method $p=2$ and $p=3$, and we compare to matching finite-volume simulations with piece-wise linear reconstruction as conventional base-line results. Our simulation sequence starts with Mach number $\mathcal{M}=0.1$, and then we modify the driving strength and the time correlation parameters of the driving routine systematically to create a sequence of simulations in which the Mach number doubles in each step, until we reach $\mathcal{M}\simeq 12.8$. The domain size for all simulations is $L=1$, and Table~\ref{tab:table_of_simulations} provides an overview of our simulation suite.

\begin{table}
    \centering
	\caption{Overview of our simulation suite and their main parameters. All boxes have their domain size fixed at $L = 1$. The $256^3$ simulations with increasing mach number are used to study the subsonic to supersonic transition. The $512^3$ and $1024^3$ are used to compare our DG method with classic finite volume with linear slope reconstruction.}
	\label{tab:table_of_simulations}
    \begin{tabular}{cccr}
    \hline
    N$^3$     & Method &   Order & ${\cal M}$    \\ \hline
    256$^3$   & DG     &   2, 3  & 0.1           \\ 
              &        &         & 0.2           \\ 
              &        &         & 0.4           \\ 
              &        &         & 0.8           \\ 
              &        &         & 1.6           \\ 
              &        &         & 3.2           \\ 
              &        &         & 6.4           \\ 
              &        &         & 12.8          \\ \hline
    512$^3$   & DG, FV &   2     & 3.2           \\
    1024$^3$  &        &         & \\
    \hline
   \end{tabular}
\end{table}

We note that in the subsonic regime, the Mach numbers realized by this procedure accurately match the expected doubling in each step, while for $\mathcal{M}$ significantly above unity, they start to fall slightly short. This is of course expected at some level due to the stronger dissipation in the supersonic case already in the driving regime. This could be corrected for by a correspondingly stronger increase in the driving strength in the supersonic regime, something that we however found not really necessary yet over the limited range in Mach numbers explored here.
Note that in each case we simulate for the same number of eddy turn-over times, which however corresponds to different absolute timespans.

\begin{figure}
\begin{center}
    \includegraphics[width=85mm]{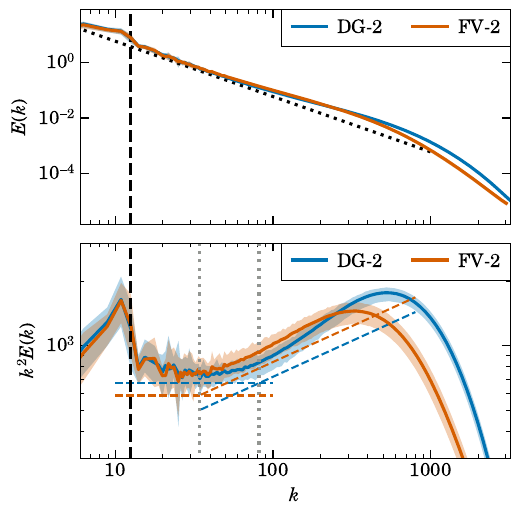}
    \end{center}
    \caption{
Velocity power spectrum of the turbulence simulation shown in Fig.~\ref{fig:compensated_structure_function}, i.e.~for a DG run with $p=2$ and a FV run, but using $1024^3$ cells with Mach number ${\cal M}=3.2$. The lines show the average of 96 measurements over five eddy turnover times and the shaded regions indicate the $1\sigma$ deviation. The top panel shows $E(k)$ directly, whereas the bottom panel displays the same data again, but this time compensated by a factor $k^2$ to compress the vertical dynamic range and highlight subtle changes in shape. The dashed vertical line marks the end of our driving range, which can be discerned as a region of elevated power. At slightly larger $k$ than this injection scale, a region with a fully developed supersonic turbulent cascade develops. This is indicated by the dashed horizontal line in the bottom panel, which has the $E(k)\propto k^{-2}$ slope of Burgers turbulence. At still smaller scales, the spectrum becomes flatter again, close to the $E(k)\propto k^{-5/3}$ expected for Kolmogorov turbulence. We have indicated this slope as an inclined dashed line in the bottom panel, with the dotted line marking the scale where extrapolations of the two power laws intersect. This intersection is reasonably close to the sonic scale inferred from the velocity structure function. We also note that there is a prominent bottleneck effect (as expected) with a small shoulder in the power spectrum before $E(k)$ drops rapidly in the dissipative regime.}
    \label{fig:sonic_point_power_spec}
\end{figure}

\begin{figure}
\begin{center}
    \includegraphics[width=85mm]{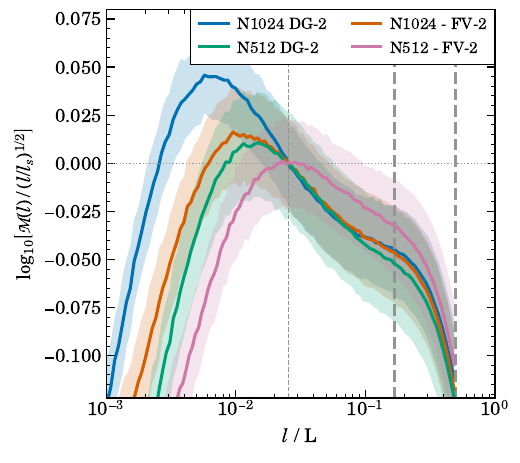}
    \end{center}
    \caption{Convergence of the velocity structure function (shown in compensated form as in the bottom panel of Fig.~\ref{fig:compensated_structure_function}) between calculations that use $512^3$ or $1024^3$ cells, and either finite volume (FV) or DG with order $p=2$, with Mach number ${\cal M}=3.2$. The lines show the average of 96 measurements over five eddy turnover times and the shaded regions indicate the $1\sigma$ deviation. The sonic scale used for rescaling the plots is the same for all lines and corresponds to the value measured for the DG run at the $1024^3$ resolution. Interestingly, the $512^3$ simulation with DG does nearly as well as the $1024^3$ run with FV, but both show at most a very feeble hint for a transition between the supersonic and subsonic regimes of turbulence. This is because of the closeness of the dissipation regime at this resolution, which already affects the region around the sonic scale strongly. For the $512^3$ run with FV, the dynamic range is clearly insufficient to resolve the region around the sonic point properly. In contrast, the high-resolution DG run is already able to distinguish different slopes of the cascade in the supersonic and supersonic regimes, although it is clear that also this calculation can still be expected to be influenced by resolution effects in the transition region.
    }
    \label{fig:sonic_point_convergence}
\end{figure}

In Figure~\ref{fig:slice_automatic_vlim}, we visually illustrate the turbulent velocity field for the ${\cal M}=0.1,\ 0.4,\ 1.6,$ and $6.4$ cases. In each panel, we show the corresponding simulations after the same number of eddy turn-over times after the start of the simulations. While the two subsonic simulations look qualitatively very similar, with the most important difference being the amplitude of the velocity field, the character of the turbulent field clearly starts to differ as we transition into the supersonic regime. This is accompanied by the appearance of strong velocity discontinuities (i.e.~shocks), and the velocity field begins to exhibit sharper gradients as well.

This difference also becomes readily apparent in a quantitative way when we consider the velocity power spectra of this sequence of simulations, which we show in Figure~\ref{fig:different_mach_numbers_vel_spectra}. We here compare the simulations with the finite volume approach (solid lines) to corresponding runs carried out with DG at orders $p=2$ (dashed) and $p=3$ (dotted). All Mach numbers from $\mathcal{M}=0.1$ to $\mathcal{M}=12.8$ are shown in a single diagram, which is readily possible as they are offset vertically due to their systematically different velocity amplitudes. The common plot makes it evident that the shape of the power spectra systematically changes when transitioning into the supersonic regime. While the inertial range outside the driving range is small due to the limited dynamic range of these simulations, it is still sufficient to show a transition from a $E(k) \propto k^{-5/3}$ Kolmogorov spectrum in the subsonic cases to a $E(k) \propto k^{-2}$ Burgers spectrum in the supersonic realizations. This trend is reproduced consistently both by the DG simulations and the finite volume scheme.

Another important and interesting trend is seen for the relative difference between the DG and the FV simulations. At given cell resolution, going from FV to DG  significantly extends the dynamic range over which the turbulence can be followed. This is already the case for $p=2$, and even more so for $p=3$, with the latter showing also signs of a more pronounced bottleneck effect, which reflects the different and generally lower numerical dissipation in this scheme. The detailed dissipation processes are also the reason why some of the DG runs show slightly enhanced velocity power again on the smallest scales, within cells. As this happens deep in the dissipation regime anyway, it is however not of concern for the practical applicability of the DG method.

Importantly, the improvement in the dynamic range brought about by $p=2$ and $p=3$ DG in the subsonic regime nearly disappears in the supersonic regime. Clearly, the accuracy advantages of DG do not play out as effectively any more for supersonic turbulence, if at all. Of course, this does not come as a complete surprise in light of our earlier discussion about the challenges involved in capturing true discontinuities with high-order DG methods. In fact, based on this we can already view it as a success that DG can robustly treat supersonic turbulence after all, with an accuracy that is at least comparable or even slightly better than that of a finite volume method. Since any supersonic cascade will eventually transition into the subsonic regime again, this is ultimately encouraging, because it means that in simulations that offer sufficiently high dynamic range, the advantages of DG can still become important again in the subsonic regime. We will explicitly return on this point in Section~\ref{sec:sonicpoint}.

In Figure~\ref{fig:different_mach_numbers_vel_structure_function}, we consider the velocity structure functions of the same sequence of Mach numbers, here shown only for DG simulations of order $k=2$ with $256^3$ cells. While the subsonic runs exhibit a scaling close to $v(l)\sim l^{1/3}$ on large scales, the structure functions noticeably steepen in the supersonic runs, reaching a scaling close to $v(l)\sim l^{1/2}$ there. This is consistent with our findings for the power spectra, and reflects the theoretical expectations for the transition from Kolmogorov to Burgers turbulence.
 
Besides considering the velocity statistics, it is also interesting to look for systematic differences in the density probability distribution function (PDF) of the turbulence. In Figure~\ref{fig:densityPDF}, we show the volume-weighted density probability distribution functions obtained by considering the density fields in different high-resolution 2D slices through the simulation boxes at different times, and then averaging the results. We compute the histograms in terms of logarithmic density, i.e.~a strictly parabolic shape of our measured density PDF would therefore correspond to a lognormal distribution. We show results for $256^3$ cell simulations with the DG method at $p=2$ and $p=3$, as well as for FV, and in each case for Mach numbers $0.8$, $1.6$, $3.2$, $6.4$ and $12.8$. The density PDFs are close to log-normal, with their width being a strong growing function of Mach number. The different numerical schemes produce results that are nearly indistinguishable at this resolution. However, close inspection of higher resolution results, for example the $1024^3$ runs of DG-1 and FV for Mach number ${\cal M} = 3.2$ shown in the lower right panel of Fig.~\ref{fig:densityPDF}, does show a small systematic difference between FV and DG after all. The latter is able to resolve slightly higher densities, reflecting its lower numerical dissipation. Importantly, this property is preserved despite our use of artificial viscosity in DG.

\begin{figure*}
    \includegraphics[width=150mm]{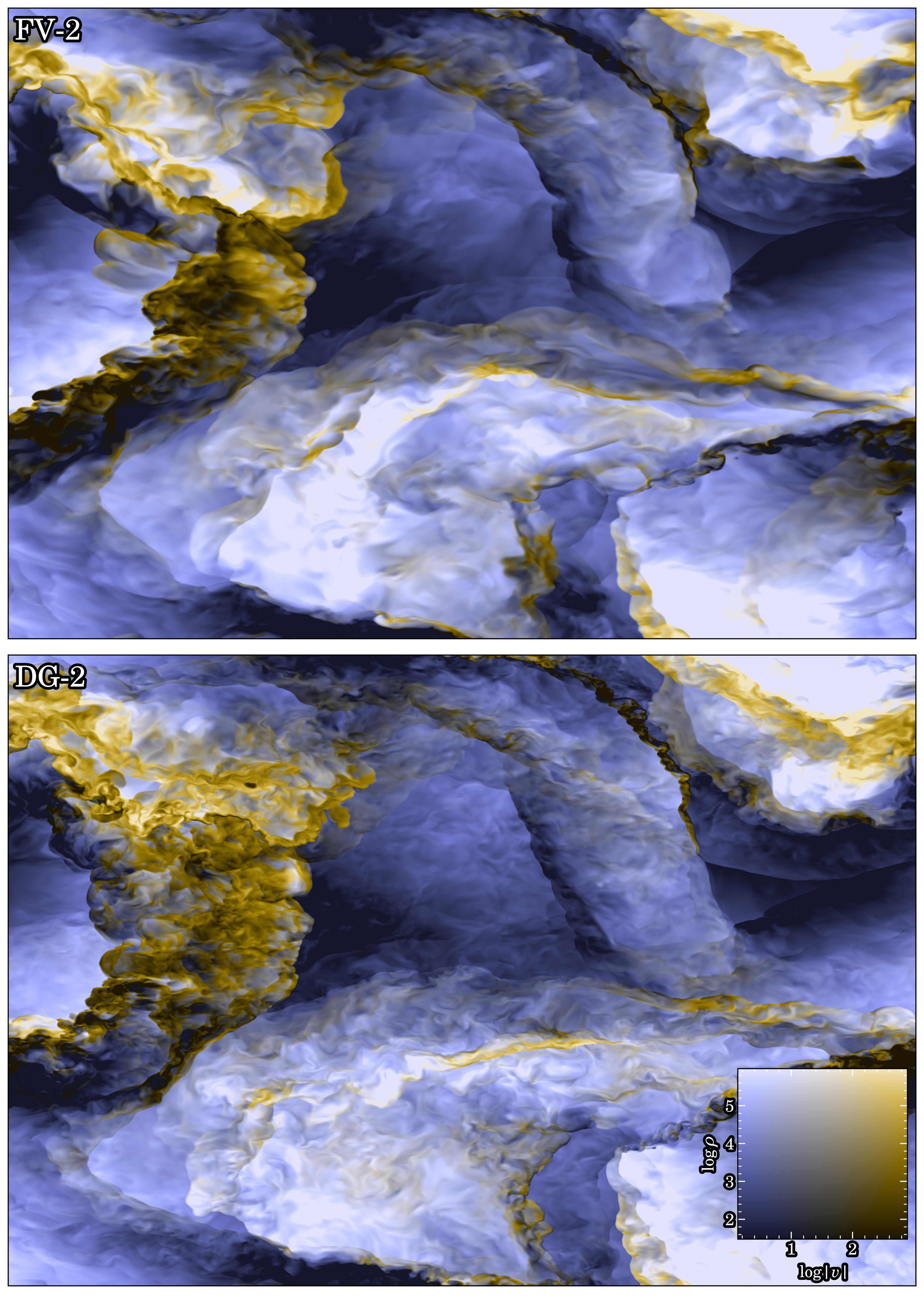}
    \caption{Visualization of the turbulence for second-order FV method with linear slope reconstruction (top panel) and DG with $p=2$ (bottom panel) simulations at Mach number ${\cal M}=3.2$. We use a two-dimensional color map, where the logarithm of density is mapped to brightness while the logarithm of the gas velocity is mapped to color hue, as indicated. The fields are shown at the same time, using $1024^3$ cells. Superficially the images look quite similar, but closer inspection reveals a richer and more pronounced small-scale structure in the DG simulation.}
    \label{fig:images_supersonic}
\end{figure*}

\section{Simulating the super- to subsonic transition}
\label{sec:sonicpoint}

As we have just seen, DG simulations offer comparatively little accuracy gains in the supersonic regime of turbulence, but they can yield considerably more accurate results in the subsonic regime by moving the numerical dissipation scale in smooth flows to smaller scales. This raises the question whether DG can still be advantageous in a simulation of supersonic turbulence when it has enough dynamic range to transition into the subsonic regime. Then we may expect that perhaps both, the transition around the so-called sonic point, as well as the subsonic part of the turbulent cascade, are represented better by the DG approach compared with traditional FV methods.

To examine this question we consider two reasonably high resolution simulations of supersonic ${\cal M}=3.2$ turbulence, carried out with $1024^3$ cells using DG at $p=2$ order, and with a piece-wise linear finite volume approach, for comparison. This resolution is  a far cry from the large $10048^3$ simulation recently used by \citet{Federrath2021}  to resolve the location of the sonic point. Since we employ a somewhat smaller Mach number to begin with, we should in principle have, however, a chance to see something if the numerical technique is less dissipative than standard finite volume approaches, given that according to Eqn.~(\ref{loc_sonic_point}) the sonic point should  be for ${\cal M}=3.2$ only  about a factor of 10 away from the injection scale. 

In Figure~\ref{fig:compensated_structure_function} we show the velocity structure function of the $p=2$ DG simulation, with the top panel directly showing the measured Mach number as a function of scale as defined in Eqn.~(\ref{eqnstrfnc}), while the bottom panel shows the structure function in a compensated way by dividing it with a $l^{1/2}$ dependence, which is the expected scaling in the regime of a self-similar supersonic turbulence cascade. Thanks to a large compression of the vertical scale, this compensated version brings out a number of important details in the shape of the structure function. In particular, on large scales, we see a settling region that ranges between the driving scale at $L/2$ down to about $\sim L/6$. Only at still smaller scales, the supersonic cascade has fully developed. Interestingly, this then follows quite accurately a \mbox{$v(l)\propto l^{0.5}$} slope until there is a quite sudden change in slope to \mbox{$v(l)\propto l^{0.4}$}, which is reasonably close to the expected slope in the region of the subsonic cascade. We thus think that this clear kink in the structure function identifies the sonic point -- the supersonic to subsonic transition. We have also identified a scale $l_s$ in the structure function where $v(l_s) = c_s$. This scale lies close to the place where we detect the change in slope in the structure function, but it is not identical to it in our results. Towards still smaller scales, the compensated structure function eventually reaches a maximum after following the \mbox{$v(l)\propto l^{0.4}$} inertial range for a while, and then declines rapidly due to a steeping of the slope.

A similar behaviour can be inferred from the velocity power spectrum, which we show in Figure~\ref{fig:sonic_point_power_spec} for the same simulation. Again, we show in the top panel the plain power spectrum, while in the bottom panel we plot it in a compensated form where the power spectrum has been multiplied with a $k^2$ factor. The latter is the expected slope for Burgers turbulence in a supersonic cascade. Indeed, our results for the velocity power spectrum accurately follow this slope over a small dynamic range beginning slightly out of the driving region, once the turbulence had a chance to fully develop  in a self-similar fashion. Eventually the power spectrum flattens to the $k^{-5/3}$ slope of Kolmogorov turbulence, which manifests as a rise in the compensated version of the plot. There is thus a clear kink in the spectrum which we can again interpret as a manifestation of the sonic point. The subsonic intertial range is however not very large in our simulation due to its limited numerical resolution, and thus the bottleneck effect from the onset of the dissipation range influences a good part of it. Towards still smaller scales, the power spectrum eventually transitions fully into begin dominated by dissipation due to the numerical viscosity of the discretization method.

It is interesting to compare the shape of the structure function between DG and a finite volume scheme, something we present in Figure~\ref{fig:sonic_point_convergence}. There we show both the $p=2$ DG simulation as well as the corresponding FV simulation for $1024^3$ cells in the same plot. In addition, we also include two further simulations computed instead with a lower $512^3$ resolution. Interestingly, while the FV simulation with $1024^3$ cells also shows a kink in the structure function where the sonic point is expected, the supersonic slope is flatter than expected, and the feature appears somewhat washed out compared to the $p=2$ DG result. In addition, the transition to the dissipation regime appears considerably earlier. It is thus evident that the DG approach represents the sonic point much more accurately than the FV scheme, and it is also much less diffusive on small scales, allowing it to capture a larger part of the subsonic turbulent cascade.

Remarkably, the $512^3$ result for $p=2$ DG is nearly as good as the $1024^3$ FV result, and it likewise shows evidence for the sonic point, albeit not as cleanly as the $1024^3$ DG result. In contrast, the $512^3$ finite volume result fails to yield any trace of the sonic point. Apparently it is simply already too diffusive. Our comparison thus confirms that DG can offer advantages even in simulations of supersonic turbulence, provided the resolution is high enough so that the transition to subsonic turbulence can be resolved and the associated subsonic cascade is of interest for the analysis of the simulation.

Finally, it is also instructive to visually verify the better small-scale resolving power of the DG approach in maps of the turbulence field. In Figure~\ref{fig:images_supersonic}, we compare visualizations of the turbulent density field in slices through the box of the $1024^3$ DG and FV runs. We show the fields at the same time, with the logarithm of the density encoded as pixel brightness while the logarithm of the gas velocity is mapped to color hue. Superficially the images look quite similar, but closer inspection reveals a richer and more pronounced small scale structure in the DG simulation. This confirms our earlier quantitative findings for the velocity structure function and the velocity power spectrum, and reflects the better resolving power of the DG approach for the subsonic part of the turbulent cascade. There is thus no question that DG simulations of supersonic turbulence deliver higher accuracy than FV simulations for an equal number of cells. But the relative computational cost of the methods is also an important aspect that needs to be considered. We will turn to this question in the next section.

\section{Discussion on computational cost}
\label{sec:discussion}

Let us now discuss the accuracy and computational cost of DG. We have shown that DG is applicable to supersonic turbulence, and that it is also fairly accurate in this regime, but apparently not significantly more so than a finite volume technique. But is it then worthwhile to go to high order given the computational cost of DG? Unlike for smooth problems, no exponential convergence can be expected in the supersonic regime, instead the width of shocks is expected to decline only as the effective spatial resolution of the scheme. Presumably this means that low-order methods are computationally more efficient if shocks play a prominent role for the evolution of the simulated system. To make this aspect more specific, we discuss in the following the expected computational cost of the DG method as a function of the employed order. This should shed some light on where potential sweet spots lie for different types of simulations.

For definiteness, we consider simulations carried out with $N^3$ cells at DG polynomial order $k$, using ideal hydrodynamics. Remember that the expected spatial convergence order $p$ of a DG method scales with the polynomal order $k$ as $p=k+1$. This means we use $b_k = \frac{1}{6}(k+1)(k+2)(k+3)$ Legendre coefficients to describe each of the $f=5$ five conserved fields (mass density, three spatial momentum densities, and energy density). The total number of degrees of freedom (DOF) is thus $D_{\rm DG} = f b_k N^3$, which is also the storage footprint of the scheme and equals the total length of the vector of weights. Note that the information content that can be captured by any simulation is arguably best given by this quantity, since the value of each degree of freedom is in principle independent of all others.

For one evaluation of the time derivative of the weights, we need to carry out an internal integration over the fluxes, as well as a flux computation over the surfaces of the cells, with the latter involving a Riemann solver. To carry out the volume integrals over the DG cells we need $v_k = (k+1)^3$ internal Gauss points each, while integrating the fluxes over each face of a cell requires $s_k = (k+1)^2$ Gauss points.

Let us first consider the volume integrations and try to estimate the number of floating point operations required for this. We shall treat them all as equivalent in cost (disregarding that divisions are more expensive), for simplicity, and will be content with an approximate count. Optimized code implementations may perhaps reach a slightly smaller operation count, for example through clever reuse of partial results, but should not be able to change the overall scaling. At each internal Gauss point, we first of all need an evolution of the field expansion to get the conserved variables of the fluid at the corresponding coordinate, which requires of order $2\,b_k\,f$ floating point operations. Conversion of the conserved states to the full hydrodynamical  flux then requires of order $c_F\simeq 15$ operations at each Gauss point. This flux is then contracted with spatial derivatives of the Legendre basis functions to give a contribution to the time derivative of the weights, the cost of this is about $6\, f b_k$ per Gauss point. Multiplying each of these partial contributions with a Gaussian quadrature weight  and then adding them up  to yield the overall contribution to the time derivative of the weights gives another $2\,f b_k$ operations per Gauss point. Summing this up for all internal Gauss points, we thus need about $v_k (10\, b_k\, f + c_F)$ operations for the internal volume integration.

For the surface fluxes, we need $6\, s_k$ Gauss points in total per cell. The evolution of the field expansions costs again $2\,b_k\,f$ operations per Gauss point. These fluid states in conserved variables are then converted to primitive variables, and are fed together with the state from the neighbouring cell to a Riemann solver, yielding the flux vector. Computing this costs at least $\sim 2\,c_F$ for a (highly) approximate Riemann solver. But since this cost has to be spent only once per interface for the two adjacent cells and Gauss points, we can approximate the cost of this part of the calculation again with $\simeq c_F$ operations per Gauss point. Contracting the fluxes with a surface normal does not require a full scalar product in this case since we know that exactly one component of the normal vector is unity, but we still need to multiply with a Legendre function value, a Gaussian quadrature weight, and add things up to the total partial result for the weight change, requiring about $3\, f b_k$ per surface Gauss point, yielding a total cost of $6\, s_k (5\,b_k\,f + c_F)$ per cell.

\begin{figure}
    \includegraphics[width=88mm]{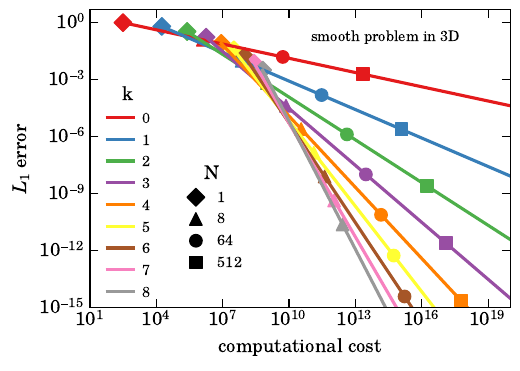}
    \caption{Expected scaling of the total numerical error as a function of the invested computational effort for a {\it smooth} hydrodynamical problem simulated with  DG at different order $k$ (coloured solid lines) in a three-dimensional box with $N^3$ cells, based on Eqn.~(\ref{eqn:totcost}).  A few illustrative problem sizes are marked with symbols, as labelled. High-order methods incur a substantially higher computational cost for a given number of cells, but they are also able to approximate a smooth solution more accurately. The error drops progressively faster as a function of resolution for higher order methods, in fact so fast that they become the method of choice -- in the sense of requiring the lowest computational cost -- for large enough problem sizes and sufficiently small target error. }
    \label{fig:different_order}
\end{figure}

\begin{figure}
    \includegraphics[width=88mm]{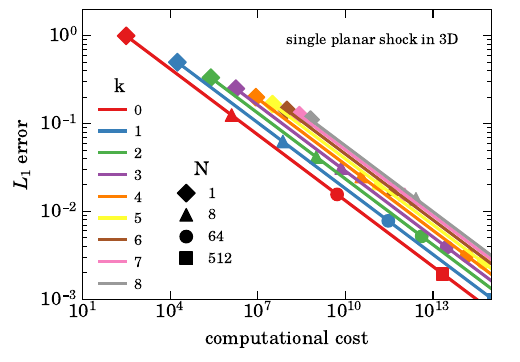}
    \caption{Expected scaling of the computational cost and total numerical error for a {\it planar shock} problem simulated with  DG at different order $k$ (coloured solid lines) in a three-dimensional box with $N^3$ cells that otherwise exhibits a  homogeneous fluid state everywhere outside the shock.  In this situation, only the numerically broadened shock itself is contributing to the error budget, which thus declines only with the linear spatial resolution as $L_1\propto N^{-1} (k+1)^{-1}$. As a result, higher-order methods do not provide a scaling advantage of their numerical error, i.e.~the relative accuracy of low and high order methods stays invariant as a function of resolution, and the higher baseline computational cost per cell of high order methods (compare the illustrative problem sizes marked with symbols) is not worthwhile. Note, however, that problems of practical interest do not consist of shocks only, rather they also have non-trivial smooth regions in between shocks, where the considerations of Fig.~\ref{fig:different_order} apply. In general, which order is computationally most cost efficient is therefore  problem-dependent.}
    \label{fig:different_order_shock}
\end{figure}

The cost of the computation of the time derivative of the DOFs (i.e.~the vector of weights, ${\rm d}\boldsymbol{w}/{\rm d}t$), is thus approximately
\begin{align}
  C_{\boldsymbol{\dot{w}}} & =   N^3 [  v_k (10\, b_k\, f + c_F) +  6 s_k (5\,b_k\,f + c_F) ]\\
              &  =  \frac{5}{3}N^3 (k+1)^2 (5k^4 + 50 k^3 + 175 k^2 + 259 k + 183), \nonumber
\end{align}
which reveals a steep $ C_{\boldsymbol{\dot{w}}} / N^3 \propto k^6$ increase of the computational cost per cell when going to high-order. Furthermore, note that the time integration itself requires several evaluations of this quantity if one aims for a consistently high order of the time integration. For the stability-preserving Runge-Kutta schemes we use, we need, for example, 2 stages for a second-order accurate scheme ($k=1$), and 3-stages for a third order scheme. For a fourth order scheme, one would ideally like to use a 4-stage Runge-Kutta scheme, which however does not exist in a purely forward form, so we need to use a 5-stage scheme instead. At still higher order, similar compromises may need to be made but we ignore some of these details here by estimating that we need $(k+1)$ evaluations of the time derivative of the DOFs to complete one timestep. Staging intermediate results according to the Butcher tableau of the Runge-Kutta scheme, and adding up the time derivatives with the Butcher weights to yield the final result at the end of the step, needs about $(k+1)^2 \times D_{\rm DG}$ floating point operations, so that we end up with a cost per timestep of about
\begin{equation}
  C_{\rm step}  =   (k+1)  C_{\boldsymbol{\dot{w}}} + (k+1)^2  f b_k N^3.
\end{equation}
Finally, the permissable Courant timestep also becomes smaller at higher order, as $\Delta t \propto h/(k+1)$, with $h \propto 1/N$. The total number of timesteps  needed  to evolve a system over some finite time interval $T$ therefore  scales in proportion to $(1+k)N$. The resulting total computational cost thus scales as
\begin{align}
  C_{\rm tot} & =   (1+k) N \left[   (k+1)  C_{\dot{w}} + (k+1)^2  f b_k N^3 \right]  \label{eqn:totcost}\\
                    & =  \frac{5}{6}N^4 (1+k)^4 ( 10k^4 + 100k^3 + 315k^2 + 523 k + 372).       \nonumber
\end{align}
To leading order, the computational cost therefore scales as $C_{\rm tot} \propto N^4 k^8$ in 3D applications of DG. This makes going to very high-order impractical, and moderate order is only worthwhile if this indeed delivers a correspondingly high accuracy.

To get a better idea of the critical trade off between computational cost and reached accuracy, we consider a fiducial error norm $L_1= a_k N^{-(k+1)}$ as a function of computational cost for different orders $k$ of the DG-scheme. A decline of the error as $L_1 \propto N^{-(k+1)}$ is expected for smooth problems, and reflects the decrease of the error norm with spatial resolution at fixed order, while for fixed resolution, the error declines exponentially with order $p=k+1$. Empirically, the coefficient $a_k$ in front typically shows only a weak dependence on order. Instead of simply taking it to be constant, we here set $a_0 =1$, and for order $k>1$ we assume that $L_1$ for $N=1$ and order $k$ is equal to the error norm for $k-1$ at the same number of degrees of freedom. In other words, we assume that for $N\simeq 1$ trading in spatial degrees of freedom for expansion order coefficients keeps the accuracy roughly fixed, which is reasonable \citep[in fact, for smooth problems, we typically expect that the accuracy improves in this case for the high-order scheme, see Fig.~15 in][]{subsonic_dg_gpu}.

In Figure~\ref{fig:different_order} we now show the expected $L_1$ error norm based on this prescription as a function of computational cost, for different order. Note that increasing the cost implicitly means considering larger problem sizes $N^3$. Despite the steep increase of the cost with order, for large problem sizes the high-order schemes tend to be advantageous, i.e.~they deliver higher accuracy at a given computational cost, or conversely, they can reach a given computational $L_1$ error for lower computational cost. For intermediate problem sizes, or low target accuracies, the situation is less clear cut, and here intermediate or lower order can be computationally advantageous. This is also confirmed by experimental findings, as reported for example in \citet{Schaal2015} and \citet{2016arXiv160209079B}. We thus conclude again that for smooth problems, where  $L_1 \propto N^{-(k+1)}$ holds, high-order DG methods tend to be cost effective. Importantly, they also have other attractive properties, such as much lower advection errors and excellent angular momentum conservation.

However, an important flipside to this discussion, which is relevant for the present paper, is that once the error norm declines more weakly with spatial resolution due to the presence of true physical discontinuities such as shocks, this cost advantage of  DG methods is defeated. We can make this more explicit by considering a fiducial simulation with a single planar shock wave in a 3D box with an otherwise homogenous fluid state. This could be realized, for example, by inflow and outflow boundary conditions on opposite sides of the box, and by matching the inflow and outflow states with the Rankine-Hugoniot jump conditions of a strong shock. In such a situation we expect the error norm to decline as $L_1 = b_k N^{-1} (k+1)^{-1}$, with an approximately constant prefactor (in the following we set $b_k =1$). The shock does become narrower with higher spatial resolution or high order, but it does so only linearly. This profoundly alters the relation between invested computational cost and expected accuracy, as illustrated in  Figure~\ref{fig:different_order_shock} for this situation. It is now always computationally favorable to reach a desired accuracy (here equivalent to shock width) by investing into higher $N$ rather than into higher order~$k$.

This does not automatically imply that higher-order DG methods are worthless for problems involving shocks, because in non-trivial flow problems there is typically a lot of interesting structure in regions outside of shocks, and these are rendered less accurately by low order methods. Also, recall that {\it most of the volume} will always be outside of shocks, given that shocks are in principle arbitrarily thin transition layers, so the volume fraction occupied by them is very small even when taking numerical broadening into account. So since one will typically be still interested in reaching high accuracy in smooth parts of the flow, a high order DG method can often be the most efficient choice. One simply then has to pay the price that some of the computational effort is consumed for an inefficient representation of shocks. In principle, this deficiency could be overcome by an approach where one applies $h$-refinement (i.e.~increasing the grid resolution while lowering order $p=k+1$) in places with shocks, while in smooth regions one should rather use $p$-refinement (increasing order $p$ while reducing mesh resolution $h$). Whether this is readily feasible in practice is however problem dependent.

\section{Conclusions}
\label{sec:conclusions}

One of the challenges of using DG methods for simulating supersonic turbulence is the appearance of Gibbs-like phenomena, especially at higher orders. These phenomena can cause spurious oscillations and numerical instabilities in the presence of shocks. The artificial viscosity method proposed by \citet{subsonic_dg_gpu} as well as the simpler artificial viscosity parametrisation proposed in the present paper can mitigate this problem, but they alone are not sufficient to stably evolve highly supersonic flows with DG at high order. 

In such flows, multiple shocks can interact, and very steep gradients of fluid variables develop inside cells. Especially for the computation of fluid states at cell surfaces, which in essence can be viewed as a polynomial extrapolation from the Gauss points inside the cells, this can frequently lead to problems. Because our code evolves the conserved variables $\boldsymbol{u} = (\rho, \rho \boldsymbol{v}, e)$, to obtain the velocity we need to divide the polynomial expansion of momentum density by that for the density, but this can lead to unphysical values at cell surfaces in the presence of very strong field variations, particularly when the density becomes very low. Similarly for the pressure, where the kinetic energy density needs to be subtracted from the total energy density, so that the pressure becomes a complicated rational function which is not necessarily as well behaved as the polynomial basis functions. This destroys the robustness of DG for highly supersonic flow.

To solve this issue we have introduced a novel projection approach of the primitive variables. They are first evaluated based on the conserved variables at the internal Gauss points, and the resulting values are used to define a polynomial expansion of the primitive variables over the cells. This regularizes the extrapolation to cell surfaces and avoids unphysical values there. Together with our simplified artificial viscosity method  this makes it possible to successfully simulate non-trivial test problems involving multiple interacting shocks as well as driven supersonic turbulence.

\ \\
\noindent Our main findings can be summarized as follows:
\begin{itemize}
\item We have introduced a simple but effective approach to capture shocks in high-order Discontinuous Galerkin discretisations of the Euler equations of fluid dynamics. It relies on the familiar von Neumann-Richtmyer viscosity applied at the internal Gauss points of each cell. This approach works well at all orders, is robust, has no storage overhead, and allows shock-capturing with sub-cell resolution.
\item Simulations with very strong shocks become much more robust and accurate if the primitive variables at cell interfaces are not simply derived from the extrapolated conserved variables there, but rather from a polynomial expansion of the primitives themselves, which is uniquely obtained from the values they assume at the internal Gauss points.
\item Using these two methodological advances, we have succeeded to stably run supersonic turbulence simulations of isothermal gas at Mach number ~12.8 with high-order DG. This has previously been met with severe stability problems, and thus this represents in its own right an important advance for the practical applicability of DG approaches in astrophysics.
\item While DG offers significant accuracy gains over finite volume schemes in the subsonic regime (which manifests itself in an extended intertial range and reduced numerical dissipation), this advantage is progressively diminished and nearly lost in the supersonic regime. However, once the sonic point is approached, the higher accuracy of DG starts to be noticeable again, and for simulations that have enough dynamic range to also resolve parts of the subsonic cascade, DG begins to shine again.
\item We have given a simple analysis of the numerical cost of our DG implementation at order $k$, based on estimating the required number of floating point operations to carry out a  3D simulation with $N^3$ cells over a fixed timespan $T$. This cost increases rapidly with order and resolution, as $\propto N^4 k^8$. For smooth problems, the error declines so rapidly for high $k$ that it is in principle still worthwhile to employ high order. The physical discontinuity of a shock defeats this scaling, however, and here low-order is more cost efficient.

\end{itemize}

Our findings in this paper make DG fully applicable to astrophysical problems involving strong shocks and contact discontinuities. Particularly the very low advection errors and accurate angular momentum conservation of DG compared with finite volume schemes should make this method interesting for many applications, for example for problems involving multiphase gas. However, as soon as many shocks are present, it appears unlikely that high-order DG methods with $p>3$ are computationally cost effective. Rather, a sweet spot can be expected for $p=2$ or $p=3$, and it is probably worthwhile to further optimize production codes for this regime.

\section*{Acknowledgements}

We thank the anonymous referee for constructive and insightful comments that helped to improve the paper. We thank Maria Werhahn for her insight into the velocity structure function. The authors acknowledge computational support from the Center for Computational Astrophysics at the  Flatiron Institute. We also acknowledge helpful discussions with numerous MPA members. 

%%%%%%%%%%%%%%%%%%%%%%%%%%%%%%%%%%%%%%%%%%%%%%%%%%
\section*{Data Availability}

Data of specific test simulations can be obtained upon reasonable request from the corresponding author.\vspace*{-0.5cm}



\bibliographystyle{mnras}
\bibliography{references} % if your bibtex file is called example.bib

%\begin{thebibliography}{99}
%\end{thebibliography}

%%%%%%%%%%%%%%%%%%%%%%%%%%%%%%%%%%%%%%%%%%%%%%%%%%

%%%%%%%%%%%%%%%%% APPENDICES %%%%%%%%%%%%%%%%%%%%%

%%%%%%%%%%%%%%%%%%%%%%%%%%%%%%%%%%%%%%%%%%%%%%%%%%

% Don't change these lines
\bsp	% typesetting comment
\label{lastpage}
\end{document}